\documentclass[journal]{IEEEtran}
\usepackage{xcolor,soul,framed} 
\colorlet{shadecolor}{yellow}
\usepackage[pdftex]{graphicx}
\graphicspath{{../pdf/}{../jpeg/}}
\DeclareGraphicsExtensions{.pdf,.jpeg,.png}
\usepackage[cmex10]{amsmath}
\usepackage{graphicx}
\usepackage{array}
\usepackage{mdwtab}
\usepackage{eqparbox}
\usepackage{url}
\usepackage{halloweenmath}
\usepackage{comment}
\usepackage{algorithm}
\usepackage{algpseudocode}
\usepackage{amsfonts}
\usepackage{multirow}
\usepackage{tikz}
\newcommand{\cm}{\checkmark}
\usepackage{wrapfig}
\usepackage{hyperref}

\usepackage{fancyhdr,lipsum}

\fancypagestyle{mahmood}{%
   \fancyhf{} 
   
   \fancyhead[C]{You can use this material personally. Reprinting or republishing this material for the purpose of advertising or promotion, creating new collective works, reselling or redistributing to servers or lists, or using any copyrighted component in other works must adhere to IEEE policy. DOI: \href{https://doi.org/10.1109/TMC.2023.3301506}{10.1109/TMC.2023.3301506}}
}%

\makeatletter
\let\ps@IEEEtitlepagestyle\ps@mahmood
\makeatother

\begin{document}
\bstctlcite{IEEEexample:BSTcontrol}
\title{Double Deep Q-Learning-based Path Selection and Service Placement for Latency-Sensitive Beyond 5G Applications}

\author{
    \IEEEauthorblockN{
        Masoud Shokrnezhad\textsuperscript{1}, Tarik Taleb\textsuperscript{1}, and Patrizio Dazzi\textsuperscript{2}
    }\\
    \IEEEauthorblockA{
        \textit{1 Oulu University, Oulu, Finland}\\
        \textit{2 University of Pisa, Pisa, Italy}\\
        \{masoud.shokrnezhad; tarik.taleb\}@oulu.fi, patrizio.dazzi@unipi.it}
}

\maketitle

\begin{abstract}\label{Sec_Abstract}
Nowadays, as the need for capacity continues to grow, entirely novel services are emerging. A solid cloud-network integrated infrastructure is necessary to supply these services in a real-time responsive, and scalable way. Due to their diverse characteristics and limited capacity, communication and computing resources must be collaboratively managed to unleash their full potential. Although several innovative methods have been proposed to orchestrate the resources, most ignored network resources or relaxed the network as a simple graph, focusing only on cloud resources. This paper fills the gap by studying the joint problem of communication and computing resource allocation, dubbed CCRA, including function placement and assignment, traffic prioritization, and path selection considering capacity constraints and quality requirements, to minimize total cost. We formulate the problem as a non-linear programming model and propose two approaches, dubbed B\&B-CCRA and WF-CCRA, based on the Branch \& Bound and Water-Filling algorithms to solve it when the system is fully known. Then, for partially known systems, a Double Deep Q-Learning (DDQL) architecture is designed. Numerical simulations show that B\&B-CCRA optimally solves the problem, whereas WF-CCRA delivers near-optimal solutions in a substantially shorter time. Furthermore, it is demonstrated that DDQL-CCRA obtains near-optimal solutions in the absence of request-specific information.
\end{abstract}

\IEEEpeerreviewmaketitle

\section{Introduction}\label{Sec_Introduction}
Nowadays, an increase in data flow has resulted in a 1000-fold increase in network capacity, which is the primary driver of network evolution. While this demand for capacity will continue to grow, the Internet of Everything is forging a paradigm shift to new-born perceptions, bringing a range of novel services with rigorous deterministic criteria, such as connected robotics, smart healthcare, autonomous transportation, and extended reality~\cite{taleb_towards_nodate}. These services will be provisioned by establishing functional components, Virtual Network Functions (VNFs), which will generate and consume vast amounts of data that must be processed in real-time to ensure service responsiveness and scalability.

In these circumstances, a distributed cloud architecture is essential~\cite{corneo_surrounded_2021}, which could be implemented via a solid cloud-network integrated infrastructure built of distinct domains in Beyond 5G (B5G)~\cite{yang_urllc_2021}. These domains can be distinguished by the technology employed, including radio access, transport, and core networks, as well as edge, access, aggregation, regional, and central clouds. Moreover, these resources can be virtualized using technologies such as Network Function Virtualization (NFV), which enables the construction of separate virtual entities on top of this physical infrastructure \cite{taleb_multi-domain_2019, taleb_cdn_2020}. Since distributed cloud and network domains would be diverse in terms of characteristics but limited in terms of capability, communication and computing resources should be jointly allocated, prioritized, and scheduled to ensure maximum Quality of Service (QoS) satisfaction while maximizing resource sharing and maintaining a deterministic system state, resulting in energy savings as one of the most significant examples of cost minimization objectives~\cite{li_cognitive_2021}.

The joint problem of resource allocation in cloud-network integrated infrastructures has been extensively studied in the literature. Emu~\textit{et al.}~\cite{emu_latency_2020} analyzed the VNF placement problem as an Integer Linear Programming (ILP) model that guarantees low End-to-End (E2E) latency while preserving QoS requirements by not exceeding an acceptable latency violation limit. They proposed an approach based on neural networks and demonstrated that it can result in near-optimal solutions in a timely way. Vasilakos~\textit{et al.}~\cite{vasilakos_towards_2021} examined the same problem and proposed a hierarchical Reinforcement Learning (RL) method with local prediction modules as well as a global learning component. They demonstrated that their method significantly outperforms conventional approaches. Sami~\textit{et al.}~\cite{sami_demand-driven_2021} investigated a similar topic to minimize the cost of allocations, and a Markov decision process design was provided. They claimed that the proposed method provides efficient placements. Performing cost-effective services was also investigated by Liu~\textit{et al.}~\cite{liu_cost-efficient_2023} and He~\textit{et al.}~\cite{he_leveraging_2023}. In the former, the authors considered the cost of computing and networking resources as well as the cost of using VNFs and proposed a heuristic algorithm, whereas, in the latter, they considered latency as a cost and proposed a Deep Reinforcement Learning (DRL) solution to the problem. Iwamoto~\textit{et al.}~\cite{iwamoto_optimal_2023} investigated the problem of scheduling VNF migrations in order to optimize the QoS degradation of all traffic flows and proposed a stochastic method on the basis of the load degree of VNF instances.

\begin{table*}[t!]
\caption{Literature Review.}
\centering
\begin{tabular}{|c|c|c|c|c|c|c|c|c|c|c|c|}
\hline

\multirow{3}{*}{\textbf{Paper}} 
& \multirow{3}{*}{\textbf{Objective}} 
& \multicolumn{7}{|c|}{\textbf{Constraints}} 
& \multicolumn{3}{|c|}{\textbf{Solution Approach}}
\\

\cline{3-12} 

\multicolumn{1}{|c|}{}
& \multicolumn{1}{c|}{}
& \multicolumn{1}{c|}{\multirow{2}{*}{\begin{tabular}[c]{@{}c@{}} Traffic\\ Priority \end{tabular}}} 
& \multirow{2}{*}{Routing}
& \multicolumn{1}{c|}{\multirow{2}{*}{\begin{tabular}[c]{@{}c@{}} Network\\ Capacity \end{tabular}}} 
& \multicolumn{2}{c|}{Network Latency} 
& \multicolumn{1}{c|}{\multirow{2}{*}{\begin{tabular}[c]{@{}c@{}} Compute\\ Capacity \end{tabular}}} 
& \multicolumn{1}{c|}{\multirow{2}{*}{\begin{tabular}[c]{@{}c@{}} Compute\\ Latency \end{tabular}}} 
& \multirow{2}{*}{Optimal}
& \multirow{2}{*}{Heuristic}
& \multirow{2}{*}{Learning}
\\

\cline{6-7} 

\multicolumn{1}{|c|}{}
& \multicolumn{1}{c|}{}
& \multicolumn{1}{c|}{}
& \multicolumn{1}{c|}{} 
& \multicolumn{1}{c|}{} 
& Device
& Link
& \multicolumn{1}{c|}{} 
& \multicolumn{1}{c|}{} 
& \multicolumn{1}{c|}{} 
& \multicolumn{1}{c|}{} 
& \multicolumn{1}{c|}{} 
\\

\hline

\cite{nguyen_virtual_2023} & max profit & \cm & \cm & \cm & & & \cm &  & & \cm & \\
\cite{liu_cost-efficient_2023} & min cost & & & \cm & & \cm & \cm &  & & \cm & \\
\cite{he_leveraging_2023} & max profit - cost & & & & & \cm & \cm & \cm & & & \cm \\
\cite{xuan_multi-agent_2023} & min energy & & \cm & & & \cm & \cm & \cm & & & \cm \\
\cite{miyamura_joint_2023} & min cost & & \cm & \cm & & & \cm & & & \cm & \\
\cite{iwamoto_optimal_2023} & max fairness & & & & & \cm & \cm & \cm & & \cm & \\
\cite{yang_online_nodate} & max profit - cost & & \cm & \cm & & & \cm & & & \cm & \\
\cite{emu_latency_2020} & min latency & & & \cm & & \cm & \cm & & & & \cm \\
\cite{vasilakos_towards_2021} & min latency & & & & & & \cm & \cm & & & \cm \\
\cite{sami_demand-driven_2021} & min cost & \cm & & & & & \cm & & & & \cm \\
\cite{kuo_deploying_2018} & max rate & & \cm & \cm & & & \cm & & & \cm & \\
\cite{mada_latency-aware_2020} & min cost & & & \cm & & \cm & \cm & \cm & \cm & & \\
\cite{zhang_adaptive_2019} & max rate & & & \cm & & \cm & \cm & \cm & & \cm & \\
\cite{yuan_toward_2020} & min latency + cost & & & & & \cm & \cm & & \cm & \cm & \\
\cite{gao_cost-efficient_2020} & min latency & & \cm & & & \cm & \cm & \cm & & \cm & \\
this work & min cost & \cm & \cm & \cm & \cm & \cm & \cm & \cm & \cm & \cm & \cm \\
\hline
\end{tabular} 
\label{Tab_lit}
\end{table*}

Although innovative techniques for addressing computing resource restrictions have been proposed by the above-mentioned authors, the network is solely considered as a pipeline in their studies, with no cognitive ability to the cloud domains. Nevertheless, there are additional studies in the literature that have been concentrating on communication and computing resources jointly. Kuo~\textit{et al.}~\cite{kuo_deploying_2018} studied the joint problem of VNF placement and path selection in order to better utilize the network resources, and a heuristic approach was proposed to tackle it. Mada~\textit{et al.}~\cite{mada_latency-aware_2020} and Zhang~\textit{et al.}~\cite{zhang_adaptive_2019} addressed the problem of VNF placement with the objective of maximizing the sum rate of accepted requests. Mada \textit{et al.} solved the problem by using an optimization solver, and Zhang \textit{et al.} adopted a heuristic strategy. Yuan, Tang and You~\cite{yuan_toward_2020} formulated the latency-optimal placement of functions as an ILP problem and proposed a genetic meta-heuristic algorithm to solve it. Gao~\textit{et al.}~\cite{gao_cost-efficient_2020} focused on the VNF placement and scheduling to reduce the cost of computing resources by proposing a latency-aware heuristic algorithm. Minimizing the cost of allocations was also investigated by Miyamura~\textit{et al.}~\cite{miyamura_joint_2023} and Yang~\textit{et al.}~\cite{yang_online_nodate}. They took into account traffic routing constraints and proposed heuristic approaches to address the problem. By considering energy consumption as the most significant cost associated with networking and computing resources, Xuan~\textit{et al.}~\cite{xuan_multi-agent_2023} addressed the same problem by proposing an algorithm based on a multi-agent DRL and a self-adaptation division strategy. Nguyen~\textit{et al.}~\cite{nguyen_virtual_2023} investigated the problem of VNF placement, where requests are weighted according to their priority and the goal is to maximize the total weight of services accepted for deployment on the infrastructure.

The methods presented in the cited studies are effective for resolving the resource allocation problem. However, such approaches cannot be utilized in B5G systems. Due to the stringent QoS requirements in the delay-reliability-rate space \cite{alwis_survey_2021}, the large number of concurrent services and requests, and the ever-changing dynamics of both infrastructure and end-user service usage behavior in terms of time and space, every detail of communication and computing resources must be determined and controlled in order to realize a deterministic B5G system \cite{yang_urllc_2021}. In some studies, latency-related limitations and requirements were simply ignored \cite{kuo_deploying_2018, miyamura_joint_2023, yang_online_nodate, nguyen_virtual_2023}. Despite the fact that delay is addressed in the other studies mentioned, they simplified it to be a connection feature, and queuing delay in network devices is completely eliminated. Furthermore, path selection is disregarded in some studies~\cite{mada_latency-aware_2020,zhang_adaptive_2019,yuan_toward_2020}, and cost optimization is overlooked in others \cite{zhang_adaptive_2019,yuan_toward_2020}.

This paper fills in the gap in the current literature by investigating the joint problem of allocating communication and computing resources, including VNF placement and assignment, traffic prioritization, and path selection. The problem is faced while taking into account capacity constraints and link and queuing delays, to minimize overall cost. As an extension of the work presented in \cite{shokrnezhad_near-optimal_2022}, the following are the primary contributions of this research:
\begin{itemize}
    \item Formulating the joint resource allocation problem of the cloud-network integrated infrastructure as a Mixed Integer Non-Linear Programming (MINLP) problem.
    \item Proposing a method based on the Branch \& Bound (B\&B) algorithm to discover the optimal solution of the problem, and devising a heuristic approach based on the Water-Filling (WF) algorithm in order to identify near-optimal solutions to the problem. When the system is fully known, both techniques can be applied to solve the problem.
    \item Developing an architecture based on the Double Deep Q-learning (DDQL) technique comprising agent design, training procedure, and decision-making strategy for allocating resources when the system is only partially known, i.e., there is no prior knowledge about the requests' requirements.
\end{itemize}

The reminder of this paper is organized as follows. Section \ref{Sec_SystemModel} introduces the system model. The resource allocation problem is formulated in Section \ref{Sec_ProblemDefinition}. Next, the B\&B and heuristic approaches are presented in Section \ref{Sec_FullyInformedMethods}. Section \ref{Sec_PartiallyInformedMethods} presents a DDQL-based resource allocation architecture. Finally, numerical results are illustrated and analyzed in Section \ref{Sec_NumericalResults}, followed by concluding remarks in Section \ref{Sec_Conclusion}.

\section{System Model}\label{Sec_SystemModel}
\begin{figure}[t!]\centering
\includegraphics[width=3in]{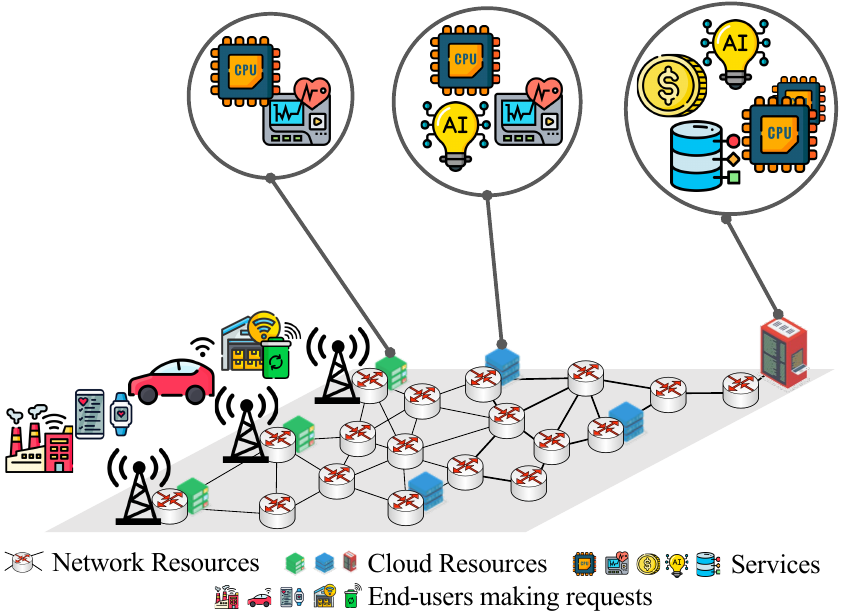}
  \caption{The envisioned system model.}
  \label{Fig_SystemModel}
\end{figure}

In the following, we describe the main components of the system envisioned in this paper. As depicted in Fig.~\ref{Fig_SystemModel}, the system consists of an infrastructure (integrated networking and computing resources), services running on computing resources, and end-user requests that must be connected to the services via networking resources. The parameters defined in this section are summarized in Table ~\ref{Tab_ParameterVariableDescription}.

\subsection{Infrastructure Model}\label{SubSec_InfrastructureModel}
The considered infrastructure is composed of the edge (non-radio side) and core network domains consisting of $\mathcal{V}$ nodes, $\mathcal{L}$ links, and $\mathcal{P}$ paths denoted by $\mathcal{G} = \langle \boldsymbol{\mathcal{V}},\boldsymbol{\mathcal{L}}, \boldsymbol{\mathcal{P}} \rangle$. $\boldsymbol{\mathcal{V}} = \{v | v \in \{1,2,...,\mathcal{V}\}\}$ is the set of nodes. $\boldsymbol{\mathcal{L}} \subset \{l:(v,v') | v,v' \in \boldsymbol{\mathcal{V}}\}$ indicates the set of links, where the bandwidth of link $l$ is constrained by $\widehat{B_l}$, and it costs $\Xi_l$ per capacity unit. Although a variety of factors (distance, technology, redundancy, accessibility, etc.) contribute to this cost as a capital expenditure, the energy used by network devices to process the traffic carried by this link is one of the significant operating expenses affecting this cost and must be precisely addressed in order to realize future networks \cite{lei_energy-saving_2021}. $\boldsymbol{\mathcal{P}} = \{p:(\vdash_p, \dashv_p) | p \subset \boldsymbol{\mathcal{L}}\}$ denotes the set of all paths in the network, where $\vdash_p$ and $\dashv_p$ are the head and tail nodes of path $p$, and $\delta_{p,l}$ is a binary constant equal to $1$ if path $p$ contains link $l$. It should be noted that all paths are directed vectors of nodes with no loops.

\begin{figure}[t!]
  \centering
  \includegraphics[width=3in]{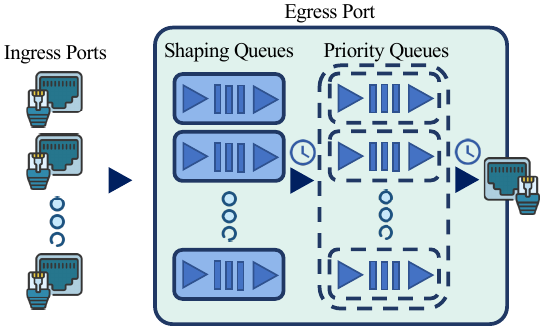}
  \caption{ATS switch model.}\label{Fig_SwitchModel}
\end{figure}

Each node in the network is an IEEE 802.1 Time-Sensitive Networking (TSN) device comprising an IEEE 802.1 Qcr Asynchronous Traffic Shaper (ATS) at each egress port. As depicted in Fig.~\ref{Fig_SwitchModel}, An ATS uses a two-level queuing model \cite{specht_urgency-based_2016}: 1) an array of shaped queues, each associated with a priority level and an ingress port, and 2) one queue per priority level. Each priority queue combines the output of all shaped queues with the same priority level. All queues implement the First-In-First-Out (FIFO) strategy. The next packet for transmission is identified by comparison of 1) the associated priority levels, and 2) the eligibility times of the Head-of-Queue (HoQ) packets. This could be accomplished, for instance, using comparator networks or linear iteration over all queues/HoQ packets while the transmission of a previous packet is in progress. 

We consider $\boldsymbol{\mathcal{K}} = \{k | k \in \{1,2,...,\mathcal{K}\}\}$ as the set of priority levels and assume that $k_r$ is the assigned priority of the traffic associated with request $r$, and the size of the queues for priority level $k$ is the same and equal to $\widehat{\mathcal{T}_k}$. Note that lower levels have higher priorities. Moreover, each node $v$ is equipped with computing resources as one of the prospective hosts to deploy service VNFs and limited to a predefined capacity threshold $\widehat{\zeta_v}$, which costs $\Psi_v$ per capacity unit. $\Psi_v$ is an increasing function of the energy consumed by various components of computing nodes (such as the processor, memory, and storage) to process requests and required by cooling systems to maintain appropriate temperatures. This cost is one of the most significant obstacles that must be overcome to make future applications feasible \cite{arshad_utilizing_2022}.

It is worth mentioning that the network is divided into several tiers, with nodes distributed across them so that the edge nodes (the entry nodes of requests) are located in tier $0$. The higher the tier index, the greater the capacity of the associated nodes, and the lower their cost. In other words, the nodes closest to end-users (or to the nodes that serve as entry points - e.g., far edge node or in-network computing nodes \cite{kianpisheh_survey_nodate}) are provisioned with high-cost, limited-capacity computing facilities, while low-cost, high-capacity depots are deployed in the core.

\subsection{Service Model}\label{SubSec_ServiceModel}
The set of services is dubbed $\boldsymbol{\mathcal{S}}=\{s |  s \in \{1,2,...,\mathcal{S}\}\}$, where $\mathcal{S}$ indicates the number of services. If an end-user requests a service, its VNF has to be replicated in the network-embedded computing resources. Each VNF is empowered to serve more than one request, and $\widehat{\mathcal{C}_s}$ indicates the maximum capacity of each VNF of service $s$.

\subsection{Request Model}\label{SubSec_RequestModel}
The set of requests asking for services is represented by $\boldsymbol{\mathcal{R}}=\{r |  r \in \{1,2,...,\mathcal{R}\}\}$, where $\mathcal{R}$ is the number of requests. Each request $r$ arrives in the network through node $v_r$ (one of the nodes equipped with a radio access base station) and intends service $s_r$, specifying its minimum necessitated service capacity, network bandwidth, and maximum tolerable delay, indicated by $\widetilde{\mathcal{C}_r}$, $\widetilde{\mathcal{B}_r}$, and $\widetilde{\mathcal{D}_r}$, respectively. In addition, $\widetilde{\mathcal{T}_r}$ and $\widetilde{\mathcal{H}_r}$, denoting the burstiness of traffic and the largest packet size for request $r$, are also assumed to be known a priori. Utilizing historical data, along with predictive data analytics methods, is one of the viable options for obtaining such accurate and realistic statistical estimates of traffic.

\begin{table}[t!]
\caption{Parameter/Variable$^\star$ description.}
\centering
\begin{tabular}{|c|c|c|}
\hline
\# & \multicolumn{1}{c|}{\textbf{P/V$^\star$}} & \multicolumn{1}{c|}{\textbf{Description}} \\
\hline
\parbox[t]{1mm}{\multirow{23}{*}{\rotatebox[origin=c]{90}{\textbf{Section II}}}} 
& $\boldsymbol{\mathcal{V}}$ & set of nodes; $|\boldsymbol{\mathcal{V}}|=\mathcal{V}$ \\
& $\boldsymbol{\mathcal{L}}$ & set of links; $|\boldsymbol{\mathcal{L}}|=\mathcal{L}$ \\
& $\widehat{B_l}$ & bandwidth of link $l$ \\
& $\Xi_l$ & cost of link $l$ per capacity unit \\
& $\boldsymbol{\mathcal{P}}$ & set of paths; $|\boldsymbol{\mathcal{P}}|=\mathcal{P}$ \\
& $\vdash_p$ & head node of path $p$ \\
& $\dashv_p$ & tail node of path $p$ \\
& $\delta_{p,l}$ & $1$ if path $p$ contains link $l$, otherwise $0$ (binary input) \\
& $\boldsymbol{\mathcal{K}}$ & set of priority levels; $|\boldsymbol{\mathcal{K}}|=\mathcal{K}$ \\
& $k_r$ & priority of request $r$ \\
& $\widehat{\mathcal{T}_k}$ & queue size of priority $k$ \\
& $\widehat{\zeta_v}$ & computing capacity of node $v$ \\
& $\Psi_v$ & cost of node $v$ per capacity unit \\
& $\boldsymbol{\mathcal{S}}$ & set of services; $|\boldsymbol{\mathcal{S}}|=\mathcal{S}$ \\
& $\widehat{\mathcal{C}_s}$ & capacity of each VNF of service $s$ \\
& $\boldsymbol{\mathcal{R}}$ & set of requests; $|\boldsymbol{\mathcal{R}}|=\mathcal{R}$ \\
& $v_r$ & point of arrival of request $r$ \\
& $s_r$ & target service of request $r$ \\
& $\widetilde{\mathcal{C}_r}$ & minimum required service capacity of request $r$ \\
& $\widetilde{\mathcal{B}_r}$ & minimum required network bandwidth of request $r$ \\
& $\widetilde{\mathcal{D}_r}$ & maximum tolerable delay of request $r$ \\
& $\widetilde{\mathcal{T}_r}$ & traffic burstiness of request $r$ \\
& $\widetilde{\mathcal{H}_r}$ & largest packet size of request $r$ \\
\hline
\parbox[t]{1mm}{\multirow{11}{*}{\rotatebox[origin=c]{90}{\textbf{Section III}}}} 
& $g_{r,v}$ & assigned node of request $r$ (binary var.) \\
& $z_{s,v}$ & assigned node of service $s$ (binary var.) \\
& $\varrho_{r,k}$ & assigned priority of request $r$ (binary var.) \\
& $\overrightarrow{f_{r,p,k}}$ & inquiry path and priority of request $r$ (binary var.) \\
& $\overleftarrow{f_{r,p,k}}$ & response path and priority of request $r$ (binary var.) \\
& $D_{r,k,l}$ & delay of request $r$ on priority $k$ and link $l$ \\
& $D_{r, s_r}$ & computing delay of request $r$ \\
& $D_{r}$ & E2E delay of request $r$ \\
& $\boldsymbol{\mathcal{R} 1}$ & requests with a priority higher than or equal to request $r$ \\
& $\boldsymbol{\mathcal{R}_2}$ & requests with a priority lower than request $r$ \\
& $\boldsymbol{\mathcal{R}_3}$ & requests with a priority higher than request $r$ \\
\hline
\end{tabular} 
\label{Tab_ParameterVariableDescription}
\end{table}

\section{Problem Definition}\label{Sec_ProblemDefinition}
This section describes the joint problem of VNF placement and assignment, traffic prioritization, and path selection. In what follows, the constraints and objective function are formulated as a MINLP problem and the problem is stated at the end of the section. The variables and parameters defined in this section are summarized in Table ~\ref{Tab_ParameterVariableDescription}.

\subsection{VNF Placement and Assignment Constraints}\label{SubSec_VNFPlacementAndAssignmentConstraints}
To arrange VNFs, each request must be first assigned a single node to serve as its service location (C1). This assignment is acceptable if the assigned node hosts a VNF for the requested service (C2). When the requests of a specific service are assigned to a particular node, they will be handled by a shared VNF. C3 ensures that the total service capacity required by these requests does not surpass the VNF's capacity. Additionally, C4 guarantees that the computing capacity of a node is not exceeded by the VNFs placed on it. Without these two constraints, both VNFs and nodes are at risk of becoming overloaded, leading to the potential termination of VNFs and congestion of requests. Such a scenario would significantly decrease the system's reliability and availability. The problem formulation becomes as follows:
\begin{align}\label{ConSet_VNFPlacementAndAssignmentConstraints}
    &\footnotesize \sum\nolimits_{\boldsymbol{\mathcal{V}}} g_{r,v} = 1, \forall r \in  \boldsymbol{\mathcal{R}},\tag{\footnotesize C1} 
    \\
    &\footnotesize g_{r,v} \leq z_{s_r,v}, \forall r, v \in \boldsymbol{\mathcal{R}}, \boldsymbol{\mathcal{V}},\tag{\footnotesize C2} 
    \\
    &\footnotesize \sum\nolimits_{\{r | r \in  \boldsymbol{\mathcal{R}} \wedge s_r = s\}} \widetilde{\mathcal{C}_r} g_{r,v} \leq \widehat{\mathcal{C}_s}, \forall v, s \in \boldsymbol{\mathcal{V}}, \boldsymbol{\mathcal{S}},\tag{\footnotesize C3}  
    \\
    &\footnotesize \sum\nolimits_{\boldsymbol{\mathcal{S}}} \widehat{\mathcal{C}_s} z_{s,v} \leq \widehat{\zeta_v}, \forall v \in \boldsymbol{\mathcal{V}},\tag{\footnotesize C4}
\end{align}
where $g_{r,v}$ and $z_{s,v}$ are binary variables. $g_{r,v}$ is $1$ if node $v$ is selected as the service node of request $r$, and $z_{s,v}$ is $1$ if service $s$ is replicated on node $v$. 

\subsection{Traffic Prioritization and Path Selection Constraints}\label{SubSec_TrafficPrioritizationAndPathSelectionConstraints}
To direct traffic, we must first ensure that each request is assigned to exactly one priority level (C5). Then, each request's (request and reply) paths are determined (C6 and C7). For each request, a single inquiry path is chosen that starts at the request's entry node and ends at the request's VNF node. The response path follows the same logic but in reverse order. The following two constraints guarantee that the two paths are chosen on the priority level assigned to each request (C8 and C9). Finally, the constraints maintaining the maximum capacity of links and queues are enforced (C10 and C11). With C10, the sum of the required bandwidth for all requests whose inquiry or response path, or both, contains link $l$ is guaranteed to be less than or equal to the link's capacity. In C11, the capacity of queues is guaranteed in the same way for each link and each priority level. The set includes:
\begin{align}\label{ConSet_TrafficPrioritizationAndPathSelectionConstraints}
    &\footnotesize \sum\nolimits_{\boldsymbol{\mathcal{K}}} \varrho_{r,k} = 1, \forall r \in  \boldsymbol{\mathcal{R}},\tag{\footnotesize C5} 
    \\
    &\footnotesize \sum\nolimits_{\{p | p \in \boldsymbol{\mathcal{P}} \wedge \vdash_p = v_r \wedge \dashv_p = v \}, \boldsymbol{\mathcal{K}}} \overrightarrow{f_{r,p,k}} = g_{r,v}, \forall r, v \in  \boldsymbol{\mathcal{R}}, \boldsymbol{\mathcal{V}},\tag{\footnotesize C6} 
    \\
    &\footnotesize \sum\nolimits_{\{p | p \in \boldsymbol{\mathcal{P}} \wedge \vdash_p = v \wedge \dashv_p = v_r \}, \boldsymbol{\mathcal{K}}} \overleftarrow{f_{r,p,k}} = g_{r,v}, \forall r, v \in  \boldsymbol{\mathcal{R}}, \boldsymbol{\mathcal{V}},\tag{\footnotesize C7}
    \\
    &\footnotesize \sum\nolimits_{\boldsymbol{\mathcal{P}}} \overrightarrow{f_{r,p,k}} = \varrho_{r,k}, \forall r, k \in  \boldsymbol{\mathcal{R}}, \boldsymbol{\mathcal{K}},\tag{\footnotesize C8}
    \\
    &\footnotesize \sum\nolimits_{\boldsymbol{\mathcal{P}}} \overleftarrow{f_{r,p,k}} = \varrho_{r,k}, \forall r, k \in  \boldsymbol{\mathcal{R}}, \boldsymbol{\mathcal{K}},\tag{\footnotesize C9}
    \\
    &\footnotesize \sum\nolimits_{\boldsymbol{\mathcal{R}}} \widetilde{\mathcal{B}_r} \sum\nolimits_{\boldsymbol{\mathcal{P}}, \boldsymbol{\mathcal{K}}} \delta_{p,l} \cdot (\overrightarrow{f_{r,p,k}} + \overleftarrow{f_{r,p,k}}) \leq \widehat{B_{l}}, \forall l \in  \boldsymbol{\mathcal{L}}, \tag{\footnotesize C10} 
    \\
    &\footnotesize \sum\nolimits_{\boldsymbol{\mathcal{R}}} \widetilde{\mathcal{T}_r} \sum\nolimits_{\boldsymbol{\mathcal{P}}} \delta_{p,l} \cdot (\overrightarrow{f_{r,p,k}} + \overleftarrow{f_{r,p,k}}) \leq \widehat{\mathcal{T}_k}, \forall k, l \in \boldsymbol{\mathcal{K}}, \boldsymbol{\mathcal{L}},\tag{\footnotesize C11}
\end{align}
where $\varrho_{r,k}$ is a binary variable that equals 1 only when the priority level assigned to request $r$ is $k$, and $\overrightarrow{f_{r,p,k}}$ and $\overleftarrow{f_{r,p,k}}$ are binary variables that reflect the inquiry and response paths for request $r$ on priority level $k$, respectively.

\subsection{Delay Constraints}\label{SubSec_DelayConstraints}
To guarantee the minimum delay requirement of requests, the following settings should be adhered:
\begin{align}\label{ConSet_DelayConstraints}
    &\footnotesize D_{r, s_r} \widetilde{\mathcal{C}_r} = \widetilde{\mathcal{H}_r}, \forall r \in \boldsymbol{\mathcal{R}},\tag{\footnotesize C12}\\
    & \footnotesize 
    \begin{aligned}
        D_{r,k,l} = &\frac{\sum\nolimits_{\boldsymbol{\mathcal{R}_1}} \widetilde{\mathcal{T}_{r'}} + \bigwedge\nolimits_{\boldsymbol{\mathcal{R}_2}} \widetilde{\mathcal{H}_{r'}}}{\widehat{B_{l}} - \sum\nolimits_{\boldsymbol{\mathcal{R}3}} \widetilde{\mathcal{B}_{r'}}} + \frac{\widetilde{\mathcal{H}_{r}}}{\widehat{B_{l}}}, \;\forall r, k, l \in \boldsymbol{\mathcal{R}}, \boldsymbol{\mathcal{K}}, \boldsymbol{\mathcal{L}},    \end{aligned}\tag{\footnotesize C13} \\
    &\scriptsize
    \begin{aligned}
        D_{r} = &\sum\nolimits_{\boldsymbol{\mathcal{P}}, \boldsymbol{\mathcal{L}}, \boldsymbol{\mathcal{K}}} D_{r,k,l} \delta_{p,l} \cdot (\overrightarrow{f_{r,p,k}} + \overleftarrow{f_{r,p,k}}) + D_{r, s_r}, \forall r \in \boldsymbol{\mathcal{R}}\\
    \end{aligned}\tag{\footnotesize C14} \\
    &\footnotesize D_{r} \leq \widetilde{\mathcal{D}_r}, \forall r \in \boldsymbol{\mathcal{R}},\tag{\footnotesize C15} 
\end{align}
where $D_{r,k,l}$, $D_{r, s_r}$ and $D_{r}$ are continuous variables denoting the delay experienced by a given ﬂow of request $r$ associated with priority level $k$ passing through ATS-based link $l$ \cite{specht_urgency-based_2016}, its computing delay, and the corresponding E2E delay calculated as the sum of the delays on the links that comprise both paths of the request and its computing delay. Besides, $\bigwedge$ is a function which returns the max value over the given set, $\boldsymbol{\mathcal{R}_1}$ equals $\{r' | r' \in \boldsymbol{\mathcal{R}} \wedge k_{r'} \leq k \wedge \delta_{p,l}(\overrightarrow{f_{r',p, k_{r'}}} + \overleftarrow{f_{r',p, k_{r'}}}) > 0\}$, $\boldsymbol{\mathcal{R}_2}$ represents $\{r' | r' \in \boldsymbol{\mathcal{R}} \wedge k_{r'} > k \wedge \delta_{p,l}(\overrightarrow{f_{r',p,k_{r'}}} + \overleftarrow{f_{r',p,k_{r'}}}) > 0 \}$, and $\boldsymbol{\mathcal{R}_3}$ denotes $\{r' | r' \in \boldsymbol{\mathcal{R}} \wedge k_{r'} < k \wedge \delta_{p,l}(\overrightarrow{f_{r',p,k_{r'}}} + \overleftarrow{f_{r',p,k_{r'}}}) > 0\}$. These sets represent requests that share the same link as request $r$, whereas $\boldsymbol{\mathcal{R} 1}$ includes requests with a higher or equal priority, $\boldsymbol{\mathcal{R}_2}$ contains requests with a lower priority, and $\boldsymbol{\mathcal{R}_3}$ shares requests with a higher priority.

\subsection{Objective Function}\label{SubSec_ObjectiveFunction}
The objective function is to minimize the total cost of allocated computing nodes and network links, that is:
\begin{equation*}\label{Eq_OF}
    \footnotesize \sum\nolimits_{\boldsymbol{\mathcal{R}}, \boldsymbol{\mathcal{V}}} \Psi_v g_{r,v} + \sum\nolimits_{\boldsymbol{\mathcal{R}}, \boldsymbol{\mathcal{L}}} \Xi_l \sum\nolimits_{\boldsymbol{\mathcal{P}}, \boldsymbol{\mathcal{K}}} \delta_{p,l}(\overrightarrow{f_{r,p,k}} + \overleftarrow{f_{r,p,k}}), \tag{OF}
\end{equation*}
As mentioned in Section \ref{Sec_SystemModel}, this cost is directly related to the energy consumption of networking and computing elements, and its reduction is a crucial open challenge that must be carefully addressed to enable B5G systems \cite{bhat_6g_2021, giordani_toward_2020, gustavsson_implementation_2021, mahmood_machine_2021}.

\subsection{Problem}\label{SubSec_Problem}
Considering the constraints and objective function, the problem of Communication and Computing Resource Allocation (CCRA) is:
\begin{equation}\label{Eq_CCRA}
    \footnotesize \text{CCRA: } \textit{ min } \text{OF} \textit{ s.t. } \text{C1 - C15.} 
\end{equation}

\section{Fully-Informed Methods}\label{Sec_FullyInformedMethods}
In this section, the system is assumed to be fully known, i.e., the list of services and their characteristics are available, and the current state of the network and cloud resources as well as requests and their requirements are being monitored and collected on a regular basis. This could be the case of an industrial environment whereby tasks and communications among robots and devices are pre-planned \cite{yu_deep_nodate, guo_probabilistic-assured_nodate, taleb_orchestrating_2019}. Under such scenarios, the following section proposes two methods, B\&B-CCRA and WF-CCRA, to solve the problem specified by (\ref{Eq_CCRA}). Clearly, an efficient strategy for implementing these methods is to centralize their development as system orchestrator components. Then, when  end-users request access to the services, the methods can be executed, and the resulting decisions can be applied to the network and cloud resources using Software-Defined Networking (SDN) and NFV technologies.
 
\subsection{B\&B-CCRA}\label{SubSec_BBCCRA}
Suppose that C1 and C5 - C15 are eliminated from (\ref{Eq_CCRA}) and only C2 - C4 affect the problem. Given this, the problem can be reformulated as minimizing the cost of assigned nodes within the capacity constraints of nodes and VNFs, that is $min \sum\nolimits_{\boldsymbol{\mathcal{R}}, \boldsymbol{\mathcal{V}}} \Psi_v g_{r,v}$ \textit{s.t.} C2 - C4. If a new parameter denoted $\Psi'_v = \mathcal{M} - \Psi_v$, where $\mathcal{M}$ is a big positive number, is defined and substituted for $\Psi_v$, the relaxed problem can be rewritten equivalently as $max \sum\nolimits_{\boldsymbol{\mathcal{R}}, \boldsymbol{\mathcal{V}}} \Psi'_v g_{r,v}$ \textit{s.t.} C2 - C4, which is the Multi-Dimensional Knapsack (MDK) problem with at least $\mathcal{S}$ items and $\mathcal{V}$ knapsacks. Since the MDK problem is NP-hard \cite{kellerer_multidimensional_2004} and a relaxed version of our problem is as hard as this problem, it is proved that our problem is also NP-hard, and finding its optimal solution in polynomial time is mathematically intractable. One potential strategy for addressing such a problem is to restrict its solution space using the B\&B algorithm, which relaxes and solves the problem to obtain lower bounds, and then improves the bounds using mathematical cuts to reach acceptable solutions. The method is described in Algorithm \ref{Alg_BB}. In this algorithm, the solution space is discovered by maintaining an unexplored candidate list $\boldsymbol{\mathcal{N}}=\{N_t | t \geq 1\}$, where each node $N_t$ contains a problem, denoted by $\Phi_t$, and $t$ is the iteration number. This list only contains $N_1$, the root candidate, at the beginning with the primary problem to be solved. To reduce its enormous computational complexity, instead of directly applying the B\&B algorithm to CCRA, we consider its integer linear transformation as the problem of $N_1$.

CCRA comprises non-linear constraints C13 and C14. To linearize C13, the summations and max function with variable boundaries should be converted to a linear form. A simple, effective technique is to replace each term with an approximated upper bound. Since the aggregated traffic burstiness is bounded by $\widehat{\mathcal{T}_{k}}$ for each priority level $k$ in C11, $\sum_{\boldsymbol{\mathcal{R}_1}} \widetilde{\mathcal{T}_{r'}}$ can be replaced by the sum of this bound for all priority levels greater than or equal to $k$, that is $\sum_{\{k'|k' \leq k\}} \widehat{\mathcal{T}_{k'}}$. In a similar way, we define a new constraint (C13$'$) for the aggregated bandwidth allowed on priority level $k$ over link $l$, dubbed $\widehat{f_{l,k}}$, and replace the sum of allocated bandwidths with $\sum_{\{k'|k' < k\}} \widehat{f_{l,k'}}$. Besides, the maximum packet size for a particular subset of requests can be replaced by the maximum permitted packet size in the network, denoted by $\widehat{\mathcal{H}}$. Therefore, the followings define the linear transformation of C13:
\begin{align}\label{ConSet_C13}
    &\footnotesize
    \begin{aligned}
        \sum\nolimits_{\boldsymbol{\mathcal{R}}} \widetilde{\mathcal{B}_r} \sum\nolimits_{\boldsymbol{\mathcal{P}}} \delta_{p,l}(\overrightarrow{f_{r,p,k}} + \overleftarrow{f_{r,p,k}}) \leq & \widehat{f_{l,k}}, \forall k \in \boldsymbol{\mathcal{K}}, \forall l \in \boldsymbol{\mathcal{L}},
    \end{aligned}\tag{\footnotesize C13$'$}\\
    &\footnotesize
    \begin{aligned}
        \widehat{D_{k,l}} = & \frac{\sum\nolimits_{\boldsymbol{\mathcal{K}_1}} \widehat{\mathcal{T}_{k'}} + \widehat{\mathcal{H}}}{\widehat{B_{l}} - \sum_{\boldsymbol{\mathcal{K}_2}} \widehat{f_{l,k'}}} + \frac{\widehat{\mathcal{H}}}{\widehat{B_{l}}}, \forall k \in \boldsymbol{\mathcal{K}}, \forall l \in \boldsymbol{\mathcal{L}},
    \end{aligned}\tag{\footnotesize C13$''$} 
\end{align}
where $\widehat{D_{k,l}}$ is the delay upper bound on link $l$ with priority level $k$, $\boldsymbol{\mathcal{K}_1}$ is $\{k'|k' \leq k\}$, and $\boldsymbol{\mathcal{K}_2}$ is $\{k'|k' < k\}$. Since $D_{r, s_r}$ is linear, C14 can be linearized by substituting the actual delay for the upper bound derived in C13$''$, and the new constraint for E2E delay is:
\begin{align}\label{ConSet_C14}
    &\footnotesize
    \begin{aligned}
        D_{r} = \sum\nolimits_{\boldsymbol{\mathcal{P}},\boldsymbol{\mathcal{L}},\boldsymbol{\mathcal{K}}} \widehat{D_{k,l}} \delta_{p,l}(\overrightarrow{f_{r,p,k}} + \overleftarrow{f_{r,p,k}}) + D_{r, s_r}, \forall r \in \boldsymbol{\mathcal{R}}.
    \end{aligned}\tag{\footnotesize C14$'$} 
\end{align}
Given this, the linear transformation of CCRA, dubbed LiCCRA, is as follows:
\begin{equation}\label{Eq_LiCCRA}
    \footnotesize \text{LiCCRA:} \textit{ min } \text{OF} \textit{ s.t. } \text{C1 - C12, C13$'$, C13$''$, C14$'$, C15}.
\end{equation}

Now, with LiCCRA as $\Phi_1$, each iteration of the B\&B algorithm begins with the selection and removal of a candidate from the unexplored list. Then, the problem of this candidate is naturally relaxed and solved, i.e., all the integer variables in the set $\{0,1\}$ are replaced with their continuous equivalents restricted by the box constraint $[0,1]$, and the relaxed problem is solved using a Linear Programming (LP) solver to obtain the solution of the relaxed problem $(\boldsymbol{\mu}^{\star}_t,\boldsymbol{\lambda}^{\star}_t)$ and the optimal objective value $\phi^{\star}_t$, where $\boldsymbol{\mu}$ is the relaxed integer variables set, and $\boldsymbol{\lambda}$ is the set of continuous variables. Next, if all relaxed variables have integer values, the obtained objective in this iteration is considered to update the best explored integer solution. Otherwise, a variable index $j$ is selected such that $\boldsymbol{\mu}^{\star}_t[j]$ is fractional, and the feasible constraints set $\pi_t$ is divided into two parts as $\pi_t^1 = \pi_t \cap \{\boldsymbol{\mu}_t[j] \leq \big\lfloor \boldsymbol{\mu}^{\star}_t[j] \big\rfloor\}$ and $\pi_t^2 = \pi_t \cap \{\boldsymbol{\mu}_t[j] \geq \big\lceil \boldsymbol{\mu}^{\star}_t[j] \big\rceil\}$. Then, two problems are formed as $\Phi_t^1 = min \text{ OF } \textit{s.t. } \pi_t^1$ and $\Phi_t^2 = min \text{ OF } \textit{s.t. } \pi_t^2$. Now, two child nodes $N_t^1$ and $N_t^2$, whose problems are $\Phi_t^1$ and $\Phi_t^2$ respectively, are put into the unexplored list. The B\&B algorithm is iterated until $\boldsymbol{\mathcal{N}}$ is empty. 

Alternatively, we can run this algorithm until a desired solving time is reached or an acceptable objective value is acquired. The key advantage of this algorithm is that it produces at least a lower bound even when the solving time is limited. As a result, it may be used to establish baselines allowing for the evaluation of alternative approaches.

\begin{algorithm}[t!]
\caption{B\&B-CCRA.}\label{Alg_BB}
\begin{algorithmic}[1]
\State $\boldsymbol{\mathcal{N}} \gets \{N_1\}$, $\eta^\star \gets +\infty$, $t \gets 0$
\State \textbf{while} $\boldsymbol{\mathcal{N}}$ is not empty \textbf{do}
\State \textbar \textbf{} $t \gets t+1$
\State \textbar \textbf{} $N_t \gets$ selects a node form $\boldsymbol{\mathcal{N}}$
\State \textbar \textbf{} $\boldsymbol{\mathcal{N}} \gets \boldsymbol{\mathcal{N}}$\textbackslash$\{N_t\}$
\State \textbar \textbf{} $(\boldsymbol{\mu}^{\star}_t,\boldsymbol{\lambda}^{\star}_t), \phi^{\star}_t \gets$ solve the relaxed problem of $\Phi_t$
\State \textbar \textbf{ if} $\boldsymbol{\mu}^{\star}_t$ is integer for all elements \textbf{then}
\State \textbar \textbf{} \textbar \textbf{} $\eta^\star \gets min(\eta^\star,\phi^\star_t)$
\State \textbar \textbf{ else if} $\phi^{\star}_t < \eta^\star$ is preserved \textbf{then}
\State \textbar \textbf{} \textbar \textbf{} $N^1_t,N^2_t \gets$ two children of $N_t$
\State \textbar \textbf{} \textbar \textbf{} $\boldsymbol{\mathcal{N}} \gets \boldsymbol{\mathcal{N}} \cup \{N^1_t,N^2_t\}$
\end{algorithmic}
\end{algorithm}

\subsection{WF-CCRA}\label{SubSec_WFCCRA}
Since the B\&B method searches the problem's solution space for the optimal solution, its complexity can grow up to the size of the solution space in the worst case~\cite{pataki_basis_2010}. Given that the size of the solution space in CCRA (or LiCCRA) for each request is $\mathcal{V}^2\mathcal{P}^2
\mathcal{K}$ considering its integer variables, the problem's overall size is $\mathcal{R}!\mathcal{V}^2\mathcal{P}^2\mathcal{K}$, considering the number of permutations of $\mathcal{R}$ requests. Therefore, finding its optimal solution for large-scale instances using B\&B is impractical in a timely manner, and the goal of this section is to devise an efficient approach based on the WF concept in order to identify near-optimal solutions for this problem. 

The WF-CCRA method is elaborated in Algorithm \ref{ALG_WF}. The first step is to initialize the vectors of parameters and variables used in (\ref{Eq_CCRA}) (or in (\ref{Eq_LiCCRA})). Following that, two empty sets, $\boldsymbol{\mathcal{R'}}$ and $\boldsymbol{\Omega}$, are established. The former maintains the set of accepted requests, and the latter stores the feasible resource combinations for each request during its iteration. Now, the algorithm iterates through each request in $\boldsymbol{\mathcal{R}}$, starting with the one with the most stringent delay requirement, and keeps track of the feasible allocations of VNF, priority, as well as inquiry and response paths based on the constraints of (\ref{Eq_CCRA}) (or (\ref{Eq_LiCCRA})). The final steps of each iteration are to choose the allocation with the lowest cost and fix it for the request, as well as to update remaining resources and the set of pending and accepted requests. When there is no pending request, the algorithm terminates.

The complexity of the WF-CCRA algorithm is $O(\mathcal{RVKP}^2)$. Although this approach is significantly more efficient than the B\&B algorithm in terms of complexity (it can be executed within milliseconds), its complexity can be further reduced by restricting the number of valid paths between each pair of nodes to a fixed-size set of paths with the lowest costs or smallest number of links. In addition, despite the fact that this algorithm implements only one of the $\mathcal{R}!$ possible permutations (serving requests in descending order of their urgency) and it converges to a solution where the cost of allocating resources to each request is locally minimized, it is expected to provide efficient solutions in terms of accuracy as well. The reason is that since requests for the same service are of the same quality and requests for all services have stringent QoS requirements, the unique permutations do not vary significantly.

\begin{algorithm}[t!]
\caption{WF-CCRA.}\label{ALG_WF}
\begin{algorithmic}[1]
\State initialize variable and parameter vectors
\State $\boldsymbol{\mathcal{R'}} \gets \{\}$, $\boldsymbol{\Omega} \gets \{\}$
\State sort $\boldsymbol{\mathcal{R}}$ in ascending order according to $\widetilde{\mathcal{D}_r}$
\State \textbf{while} $\boldsymbol{\mathcal{R}}$ is not empty \textbf{do}
\State \textbar \textbf{ for} $v \in \boldsymbol{\mathcal{V}}$ \textbf{do}
\State \textbar \textbf{} \textbar \textbf{ if} $z_{s_r,v}==1$ \textbf{and} $\widetilde{\mathcal{C}_r} \leq \widehat{\mathcal{C}_{s_r}}$ on $v$ \textbf{then}
\State \textbar \textbf{} \textbar \textbf{} \textbar \textbf{} $g_{r,v}=1$
\State \textbar \textbf{} \textbar \textbf{ if} $z_{s_r,v} \neq 1$ \textbf{and} $\widehat{\mathcal{C}_{s_r}} \leq \widehat{\zeta_v}$ \textbf{then}
\State \textbar \textbf{} \textbar \textbf{} \textbar \textbf{} $z_{s_r,v}=1$, $g_{r,v}=1$
\State \textbar \textbf{} \textbar \textbf{ else go to the next iteration}
\State \textbar \textbf{} \textbar \textbf{ for} $k \in \boldsymbol{\mathcal{K}}$ \textbf{do}
\State \textbar \textbf{} \textbar \textbf{} \textbar \textbf{} $\varrho_{r,k}=1$
\State \textbar \textbf{} \textbar \textbf{} \textbar \textbf{ for} $p \in \boldsymbol{\mathcal{P}} \wedge \vdash_p = v_r \wedge \dashv_p=v$ \textbf{do}
\State \textbar \textbf{} \textbar \textbf{} \textbar \textbf{} \textbar \textbf{ if} $\widetilde{\mathcal{B}_r} \leq \widehat{\mathcal{B}_{l}} \; \& \; \widetilde{\mathcal{T}_r} \leq \widehat{\mathcal{T}_{k}}$ on $l \; \forall l \in \boldsymbol{\mathcal{L}} \wedge \delta_{p,l}=1$ \textbf{then} 
\State \textbar \textbf{} \textbar \textbf{} \textbar \textbf{} \textbar \textbf{} \textbar \textbf{} \tiny $\overrightarrow{f_{r,p,k}} = 1$ \normalsize
\State \textbar \textbf{} \textbar \textbf{} \textbar \textbf{} \textbar \textbf{} \textbar \textbf{ for} $p' \in \boldsymbol{\mathcal{P}} \wedge \vdash_{p'} = v \wedge \dashv_{p'}=v_r $ \textbf{do}
\State  \textbar \textbf{} \textbar \textbf{} \textbar \textbf{} \textbar \textbf{} \textbar \textbf{} \textbar \small \textbf{ if} $\widetilde{\mathcal{B}_r} \leq \widehat{\mathcal{B}_{l}} \; \& \; \widetilde{\mathcal{T}_r} \leq \widehat{\mathcal{T}_{k}}$ on $l \; \forall l \in \boldsymbol{\mathcal{L}} \wedge \delta_{p,l}=1$ \textbf{then} \normalsize
\State \textbar \textbf{} \textbar \textbf{} \textbar \textbf{} \textbar \textbf{} \textbar \textbf{} \textbar \textbf{} \textbar \textbf{} \tiny $\overleftarrow{f_{r,p',k}} = 1$ \normalsize
\State \textbar \textbf{} \textbar \textbf{} \textbar \textbf{} \textbar \textbf{} \textbar \textbf{} \textbar \textbf{} \textbar \textbf{} calculate $D_r$ based on (C14) (or C14$'$)
\State \textbar \textbf{} \textbar \textbf{} \textbar \textbf{} \textbar \textbf{} \textbar \textbf{} \textbar \textbf{} \textbar \textbf{ if} $D_{r} \leq \widetilde{\mathcal{D}_r}$ \textbf{then} 
\State \textbar \textbf{} \textbar \textbf{} \textbar \textbf{} \textbar \textbf{} \textbar \textbf{} \textbar \textbf{} \textbar \textbf{} \textbar \textbf{} $\boldsymbol{\Omega} \gets \boldsymbol{\Omega} \cup \{(z_{s_r,v}, g_{r,v}, \varrho_{r,k}, \overrightarrow{f_{r,p,k}}, \overleftarrow{f_{r,p',k}}) \}$ 
\State \textbar \textbf{} fix assignments of $\textit{argmin}_{\boldsymbol{\Omega}}$ OF for $r$
\State \textbar \textbf{} \small update capacities, \normalsize $\boldsymbol{\Omega} \gets \{\}$, $\boldsymbol{\mathcal{R}} \gets \boldsymbol{\mathcal{R}}/\{r\}$, $\boldsymbol{\mathcal{R'}} \gets \boldsymbol{\mathcal{R'}} \cup \{r\}$
\end{algorithmic}
\end{algorithm}

\section{Partially-Informed Method}\label{Sec_PartiallyInformedMethods}
In the previous section, we assumed that the system is fully known. In this section, we consider a scenario wherein the system is only partially known, i.e., the state of the available network and cloud resources is tracked in real-time, and the list of services and their associated characteristics have been introduced in advance. However, end-users and the orchestrator do not exchange information pertaining to requests and their requirements. In this particular scenario, to solve the problem stated in (\ref{Eq_CCRA}), we employ the DDQL technique, proposed by Google in the DeepMind project \cite{hasselt_deep_2016}. In what follows, The DDQL concept and its agent, which serves as the core building block of the learning logic, are briefly introduced. The design of the learning algorithm and the architecture of the DDQL-based resource allocation approach are then discussed, along with an analysis of various implementation strategies.

\subsection{Double Deep Q-Learning Agent}\label{SubSec_DoubleDeepQLearningAgen}
RL is a technique wherein an agent is trained to tackle sequential decision problems through trial-and-error interactions with the environment. Q-learning is a widely used RL algorithm wherein the agent learns the value of each action, defined as the sum of future rewards associated with performing that action, and then follows the optimal policy, which is choosing the action with the highest value in each state. 

According to Watkins and Dayan~\cite{watkins_technical_1992}, one method for obtaining the optimal action-value function is to define a Bellman equation as a straightforward value iteration update using the weighted average of the old value and the new information, that is
\begin{equation}\label{Eq_BellmanEquation} 
\footnotesize Q(\theta_\tau, a_\tau)\;  \mathrel{{+}{=}} \; \sigma[Y_\tau^Q - Q(\theta_\tau, a_\tau)],
\end{equation}
where $\theta_\tau$ and $a_\tau$ are the agent's state and action at time slot $\tau$ respectively, $\sigma$ is a scalar step size, and $Y_\tau^Q$ is the target, defined by
\begin{equation}\label{Eq_Target} 
\footnotesize Y_\tau^Q  = \beta_{\tau+1} + \gamma \; \text{max}_{a \in \boldsymbol{\mathcal{A}}} Q(\theta_{\tau+1}, a),
\end{equation} 
where $\beta_{\tau+1}$ is the reward at time slot $\tau+1$, $\gamma \in [0,1]$ is a discount factor that balances the importance of immediate and later rewards, and $\boldsymbol{\mathcal{A}}$ is the set of actions. Since most interesting problems are too large to discover all possible combinations of states and actions and learn all action-values, one potential alternative is to use a Deep Neural Network (DNN) to approximate the action-value function. In a Deep Q-Network (DQN), the state is given as the input and the $Q$ function of all possible actions, denoted by $Q(\theta, .; \boldsymbol{\mathcal{W}})$, is generated as the output, where $\boldsymbol{\mathcal{W}}$ is the set of DNN parameters. The target of the DQN is as follows:
\begin{equation}\label{Eq_DQNTarget} 
\footnotesize Y_\tau^{DQN}  = \beta_{\tau+1} + \gamma \; \text{max}_{a \in \boldsymbol{\mathcal{A}}} Q(\theta_{\tau+1}, a, \boldsymbol{\mathcal{W}}_\tau),
\end{equation}
and the update function of $\boldsymbol{\mathcal{W}}$ is
\begin{equation}\label{Eq_DQNBellmanEquation} 
\footnotesize \boldsymbol{\mathcal{W}}_{\tau+1} = \boldsymbol{\mathcal{W}}_{\tau} +  \sigma[Y_\tau^{DQN} - Q(\theta_\tau, a_\tau; \boldsymbol{\mathcal{W}}_\tau)] \nabla_{\boldsymbol{\mathcal{W}}_\tau} Q(\theta_\tau, a_\tau; \boldsymbol{\mathcal{W}}_\tau).
\end{equation}

\begin{figure}[t!]\centering
\includegraphics[width=3.5in]{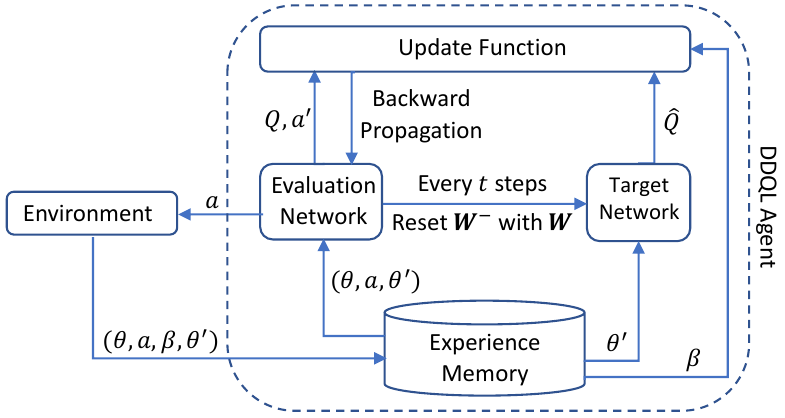}
  \caption{DDQL Agent.}
  \label{Fig_DDQLAgent}
\end{figure}

To further enhance the efficiency of DQN, it is necessary to consider two additional improvements. The first is the use of an experience memory \cite{mnih_human-level_2015}, wherein the observed transitions are stored in a memory bank, and the neural network is updated by randomly sampling from this pool. The authors demonstrated that the concept of experience memory significantly improves the DQN algorithm's performance. The second is to employ the concept of Double Deep Q-Learning (DDQL), introduced in \cite{hasselt_deep_2016}. In both standard Q-learning and DQNs, the max operator selects and evaluates actions using the same values (or the same $Q$). Consequently, overestimated values are more likely to be selected, resulting in overoptimistic value estimations. DDQL implements decoupled selection and evaluation processes. The following is the definition of the target in DDQL:
\begin{equation}\label{Eq_DDQLTarget} 
\footnotesize Y_\tau^{DDQL}  = \beta_{\tau+1} + \gamma \; \widehat{Q}(\theta_{\tau+1}, a', \boldsymbol{\mathcal{W}}^-_\tau),
\end{equation}
where $a' = argmax_{a \in \boldsymbol{\mathcal{A}}} Q(\theta_{\tau+1}, a, \boldsymbol{\mathcal{W}}_\tau)$, and the update function is
\begin{align}\label{Eq_DDQLBellmanEquation} 
\footnotesize 
\boldsymbol{\mathcal{W}}_{\tau+1} = &\boldsymbol{\mathcal{W}}_{\tau} + \sigma[Y_\tau^{DDQL} \notag \\
 & \footnotesize - Q(\theta_\tau, a_\tau; \boldsymbol{\mathcal{W}}_\tau)]\nabla_{\boldsymbol{\mathcal{W}}_\tau} Q(\theta_\tau, a_\tau; \boldsymbol{\mathcal{W}}_\tau).
\end{align}
In this model, $\boldsymbol{\mathcal{W}}$ is the set of weights for the main (or evaluation) $Q$ and is updated in each step, whereas $\boldsymbol{\mathcal{W}}^-$ is for the target $\widehat{Q}$ and is replaced with the weights of the main network every $t$ steps. In other words, $\widehat{Q}$ remains a periodic copy of $Q$. The authors demonstrated that the DDQL algorithm not only mitigates observed overestimations but also significantly improves accuracy. The training procedure of the DDQL agent is depicted in Fig. \ref{Fig_DDQLAgent}, which includes receiving the environment response and storing it in the memory bank, passing transitions to the evaluation network and updating its weights with the update function, and adjusting the weights of the target network. In this figure, $\theta'$ is the resulted state after applying action $a$.

\subsection{DDQL-CCRA}\label{SubSec_DDQLCCRA}
Since the CCRA problem comprises different sets of variables and their corresponding constraints, to solve it based on the DDQL agent depicted in Fig. \ref{Fig_DDQLAgent}, the first step is to design a chain of agents, each of which is responsible for addressing one group of the variables. Our proposed chain consists of four DDQL agents. The first agent, denoted by $\Lambda^{SP}$, is intended to determine the location of service VNFs in response to requests ($\boldsymbol{g}$ and $\boldsymbol{z}$), and thus its action set is the set of network nodes. In other words, $a^{SP} \in \boldsymbol{\mathcal{A}}^{SP} = \boldsymbol{\mathcal{V}}$. $\Lambda^{PA}$ is the second agent with action set $\boldsymbol{\mathcal{K}}$, and it is responsible for assigning the priority level of traffic. The remaining two agents route traffic by determining the inquiry path from the entering node to its VNF location and the response path in the opposite direction, denoted by $\Lambda^{QPS}$ and $\Lambda^{PPS}$, respectively. The action set of these agents comprises all possible network paths. To interact with the system, each agent provides an action that contains the index of the request for which it is attempting to satisfy its resource requirements and a value from its action space. For example, $a^{SP} = \{ r: 1, \xi: 3 \}$ means that the VNF for request $1$ should be located in node $3$, or $g_{1,3}=1$. Moreover, $\boldsymbol{a}=\{a^{SP}, a^{PA}, a^{QPS}, a^{PPS}\}$  represents the set of all agents' actions.

Next, the system state has to be formulated. As the infrastructure is the only side of the system that is known, the state is a collection of network and cloud resources, that is:
\begin{equation}\label{Eq_SystemState} 
\footnotesize \begin{split}
\theta = & \left[\widehat{\zeta_v} \forall v \in \boldsymbol{\mathcal{V}} \right] \oplus \left[\Psi_v \forall v \in \boldsymbol{\mathcal{V}} \right] \oplus \left[ \widehat{B_{l}} \forall l \in \boldsymbol{\mathcal{L}} \right] \oplus \left[ \Xi_l \forall l \in \boldsymbol{\mathcal{L}} \right] \oplus \\
& \left[ \widehat{D_{k,l}} \forall k \in \boldsymbol{\mathcal{K}}, \forall l \in \boldsymbol{\mathcal{L}} \right],
\end{split}
\end{equation}
where $\oplus$ returns the concatenation of two arrays. When the system receives actions, the state of the available network and cloud resources is updated by deactivating the resources assigned to the associated request, and resulted state $\theta'$ is generated.

The final step is to design the reward, which is a reaction to the effectiveness of action after receiving it and shifting from state $\theta$ to resulted state $\theta'$. In other words, agents are wired to the system via the reward. To address the problem defined in (\ref{Eq_CCRA}), we propose the reward as follows for request $r$:
\begin{equation}\label{Eq_Reward} 
\footnotesize \beta = 100 \left(1 - \frac{\text{OF}_{r,\boldsymbol{a}} - min \text{ OF}_{r}}{max \text{ OF}_{r} - min \text{ OF}_{r}} \right)\chi_{r, \boldsymbol{a}}
\end{equation}
where $max \text{ OF}_{r}$ and $min \text{ OF}_{r}$ are the maximum and minimum costs that can be achieved by allocating the available network and cloud resources to request $r$ without considering any constraints or requirements, $\text{OF}_{r, \boldsymbol{a}}$ is the cost of the allocations provided by the agents, and $\chi_{r, \boldsymbol{a}}$ represents the response of request $r$ to actions $\boldsymbol{a}$. $\chi_{r, \boldsymbol{a}}$ ensures that all constraints of (\ref{Eq_CCRA}) are met. Consider $\boldsymbol{a}$ containing an action that violates one of the constraints (for example, a node or a VNF or a path is overloaded, or a priority level is assigned in such a way that the E2E delay requirement is violated). In this circumstance, the affected request will respond with $\chi_{r, \boldsymbol{a}}=0$, and the reward for $\boldsymbol{a}$ will be $0$. Therefore, the probability of selecting that action decreases, and after a certain number of iterations, actions with infeasible allocations are implicitly removed from the set of possible actions. Besides, $\text{OF}_{r, \boldsymbol{a}}$ controls the efficiency of $\boldsymbol{a}$. Similarly, after a number of iterations, allocations with lower costs will have a greater chance of being selected. Therefore, after training, agents will choose feasible actions (within the constraints of (\ref{Eq_CCRA})) with lower costs (minimizing \ref{Eq_OF}).

Now, Algorithm \ref{Alg_DDQLCCRA} details DDQL-CCRA, the learning algorithm proposed to solve the CCRA problem based on DDQL. The algorithm is divided into two phases:
\subsubsection{Training Phase}\label{SubSubSec_TrainingPhase}
In this phase (lines 1 to 24), $T$ represents the number of training steps, whereas $\epsilon'$ and $\widetilde{\epsilon}$ are small positive integers to control the $\epsilon$-greedy algorithm. Through each step, the set of actions is determined and transmitted to the system, after which the reward and the updated state are received and used to train the agents employing the ADAM optimizer \cite{kingma2014adam} and update their DNN weights via the memory bank. This process is repeated over the set of requests until the specified maximum number of steps is reached. It is worth mentioning that the action in each agent is selected by an $\epsilon$-greedy policy that follows the evaluation function of the corresponding agent with probability $(1-\epsilon)$ and selects a random action with probability $\epsilon$. The probability is decreased linearly from $\epsilon$ to $\widetilde{\epsilon}$ during the training process. Using the $\epsilon$-greedy method and the ADAM optimizer ensures the convergence of DDQL-CCRA to feasible, low-cost solutions (based on the defined reward) \cite{watkins1992q}.  
\subsubsection{Decision Making Phase}\label{SubSubSec_DecisionMakingPhase}
\begin{figure}[t!]\centering
\includegraphics[width=2.8in]{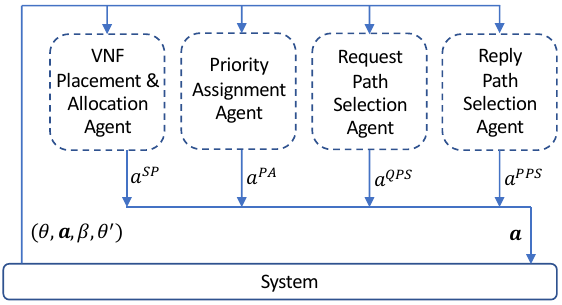}
  \caption{Data flow between the agents and the system.}
  \label{Fig_DataFLow}
\end{figure}

In each step of this phase (lines 25 to 36), one request is selected, and its required resources are allotted by the agents. The decision is then transmitted to the system to collect the infrastructure's response and the request. Following this, the reward and the mean reward, denoted by $\overline{\beta}$, are determined. Fig. \ref{Fig_DataFLow} depicts the actions generated by the agents, their transmission to the environment, and their subsequent return to the agents in preparation for the next decision-making. Due to the fact that we have no knowledge of the requests' requirements, every change in the criteria is managed by examining the average reward; if it falls below a specified threshold, denoted by $\widetilde{\beta}$, it indicates that end-users have adopted a new policy and the training phase must be repeated. This procedure continues until the required resources for each request have been determined. 

\begin{algorithm}[t!]
\caption{DDQL-CCRA.}\label{Alg_DDQLCCRA}
\begin{algorithmic}[1]
\State initialize $T$, $\epsilon'$ and $\widetilde{\epsilon}$
\State $\tau \gets 0$, $\epsilon \gets 1$, $memory \gets \{\}, \overline{\beta} \gets 0$
\State $\theta_{\tau} \gets$ get the environment's current state
\State \textbf{while} there is active request \textbf{do}
\State \textbar \textbf{} \textbf{while} $\tau \leq T$ \textbf{do}
\State \textbar \textbf{} \textbar \textbf{} $\boldsymbol{a} \gets \{\}$
\State \textbar \textbf{} \textbar \textbf{} $r \gets$ select one of the active requests
\State \textbar \textbf{} \textbar \textbf{} \textbf{for} $i$ in \{$SP, PA, QPS, PPS$\} \textbf{do}
\State \textbar \textbf{} \textbar \textbf{} \textbar \textbf{} $n \gets$ generate one random number between $0$ and $1$
\State \textbar \textbf{} \textbar \textbf{} \textbar \textbf{} \textbf{if} $n > \epsilon$ \textbf{then}
\State \textbar \textbf{} \textbar \textbf{} \textbar \textbf{} \textbar \textbf{} $\xi \gets argmax_{a \in \boldsymbol{\mathcal{A}}^{i}} Q(\theta_{\tau}, a, \boldsymbol{\mathcal{W}}_\tau)$ by $\Lambda^{i}$
\State \textbar \textbf{} \textbar \textbf{} \textbar \textbf{} \textbf{else}
\State \textbar \textbf{} \textbar \textbf{} \textbar \textbf{} \textbar \textbf{} $\xi \gets$ select a random $a$ from $\boldsymbol{\mathcal{A}}^{i}$
\State \textbar \textbf{} \textbar \textbf{} \textbar \textbf{}  $a^{i} \gets \{ r, \xi\}$
\State \textbar \textbf{} \textbar \textbf{} $\boldsymbol{a} \gets \{a^{SP}, a^{PA}, a^{QPS}, a^{PPS}\}$
\State \textbar \textbf{} \textbar \textbf{} apply $\boldsymbol{a}$
\State \textbar \textbf{} \textbar \textbf{} collect the infrastructure and the request responses
\State \textbar \textbf{} \textbar \textbf{} calculate $\beta$
\State \textbar \textbf{} \textbar \textbf{} $memory \gets memory \cup \{(\theta, \boldsymbol{a}, \beta, \theta')\}$
\State \textbar \textbf{} \textbar \textbf{} choose a sample form $memory$, and train agents
\State \textbar \textbf{} \textbar \textbf{} \textbf{if} $\epsilon > \widetilde{\epsilon}$ \textbf{then}
\State \textbar \textbf{} \textbar \textbf{} \textbar \textbf{} $\epsilon \gets \epsilon - \epsilon'$
\State \textbar \textbf{} \textbar \textbf{} $\tau \gets \tau+1$
\State \textbar \textbf{} \textbar \textbf{} $\theta_{\tau} \gets \theta'$

\State \textbar \textbf{} $\boldsymbol{a} \gets \{\}$ 
\State \textbar \textbf{} $r \gets$ select one of the active requests
\State \textbar \textbf{} \textbf{for} $i$ in \{$SP, PA, QPS, PPS$\} \textbf{do}
\State \textbar \textbf{} \textbar \textbf{} $\xi \gets argmax_{a \in \boldsymbol{\mathcal{A}}^{i}} Q(\theta_{\tau}, a, \boldsymbol{\mathcal{W}}_\tau)$ by $\Lambda^{i}$
\State \textbar \textbf{} \textbar \textbf{}  $a^{i} \gets \{ r, \xi\}$
\State \textbar \textbf{} $\boldsymbol{a} \gets \{a^{SP}, a^{PA}, a^{QPS}, a^{PPS}\}$
\State \textbar \textbf{} apply $\boldsymbol{a}$
\State \textbar \textbf{} collect the infrastructure and the request responses
\State \textbar \textbf{} calculate $\beta$
\State \textbar \textbf{} $\overline{\beta} \gets \gamma\beta + (1-\gamma)\overline{\beta}$, $\theta_{\tau} \gets \theta'$
\State \textbar \textbf{} \textbf{if} $\overline{\beta} < \widetilde{\beta}$ \textbf{then}
\State \textbar \textbf{} \textbar \textbf{} go to 1
\end{algorithmic}
\end{algorithm}

\subsection{DDQL-CCRA Resource Allocation Architecture}\label{SubSec_DDQLbasedResourceAllocationArchitecture}

\begin{figure}
  \centering
  \includegraphics[width=3in]{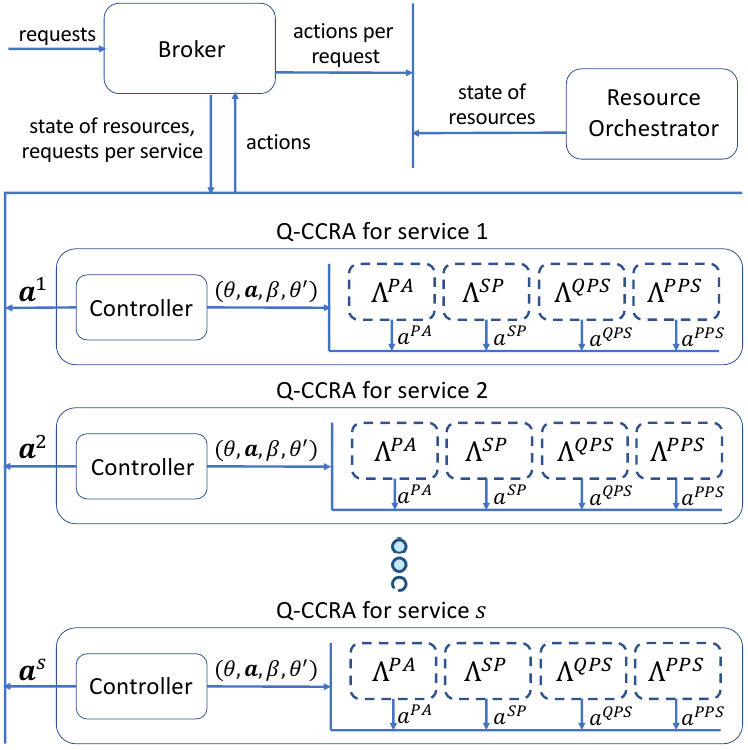}
  \caption{DDQL-CCRA Resource Allocation Architecture.}\label{Fig_DDQLbasedResourceAllocationArchitecture}
\end{figure}

The architecture of the DDQL-CCRA resource allocation method is depicted in Fig. \ref{Fig_DDQLbasedResourceAllocationArchitecture}. Due to the fact that the characteristics of different services may be entirely different, an isolated DDQL-CCRA algorithm is designed to be executed for each service. The broker receives requests, classifies them, and forwards each service's requests to its respective controller. In addition, the broker collects the most recent state of the network and cloud resources from the resource orchestrator and transmits it to the controllers. The controller is responsible for executing the DDQL-CCRA algorithm by implementing the memory bank, maintaining the state of requests, calculating the reward, and returning action sets to the broker. Action sets are collected by the broker from all controllers and relayed to the resource orchestrator to apply to the infrastructure. Since actions are chosen at random during the training phase, digital twins could be used to evaluate them to prevent the infrastructure from entering unpredictable states that result in disruptions to its operation \cite{nguyen_digital_2021}.

In order to enhance the scalability of this architecture, rather than considering the set of all nodes as the action set of $\Lambda^{SP}$ and the set of all paths of the network as the action set of $\Lambda^{QPS}$ and $\Lambda^{PPS}$, these spaces can be pruned to create fixed-size sets consisting of the most likely options for VNF placement and path selection.
\begin{itemize}
    \item For $\Lambda^{SP}$, the lower and upper boundaries of the QoS requirements for each service can be extracted (or considered inputs to the problem), and then a set with size $\mathcal{V'}$, named $\boldsymbol{\mathcal{V'}}$, including feasible nodes to maintain the QoS boundaries at the lowest cost ($\Psi_v$) is generated.
    \item For $\Lambda^{QPS}$ and $\Lambda^{PPS}$, a set of size $\mathcal{P'}$ is created for each service containing feasible paths in order to maintain the QoS boundaries at the lowest cost ($\Xi_l$). Note that these paths should begin at the edge devices (the entry nodes of requests) and terminate at one of the nodes of $\boldsymbol{\mathcal{V'}}$ for $\Lambda^{QPS}$. In $\Lambda^{PPS}$, the same logic is followed, but in reverse order.
\end{itemize}
The complexity and accuracy of the DDQL-CCRA algorithm can be modified by adjusting the size of these sets. $\mathcal{V'}$ and $\mathcal{P'}$ can be set to large numbers if high precision is required or if the complexity of running the DDQL-CCRA algorithm can be handled by high-powered software/hardware. Alternatively, small sets can be utilized to return the result in a relatively shorter amount of time.


\section{Numerical Results}\label{Sec_NumericalResults}
\begin{table}[t!]
\caption{Simulation Parameters.}
\begin{center}
\begin{tabular}{|c|c|}
\hline
\textbf{Parameter} & \textbf{Value} \\
\hline
\multirow{2}{*}{number of links ($\mathcal{L}$)} & $\sim \mathcal{U}\{3\mathcal{V}, 5\mathcal{V}\}$, where the \\ & resulted graph is connected. \\
number of priority levels ($\mathcal{K}$) & $4$ \\
number of services ($\mathcal{S}$) & $3$ \\
number of tiers & $3$ \\
capacity per link ($\widehat{B_l}$) & $\sim \mathcal{U}\{250, 300\}$ mbps \\
cost per link ($\Xi_l$) & $\sim \mathcal{U}\{10, 20\}$ \\
computing capacity per node ($\widehat{\zeta_v}$) & $\sim 100 \; \mathcal{U}(x, x+1)$ mbps \\
cost per node ($\Psi_v$) & $10^{\; x+1}$ \\
bandwidth bound per link-priority ($\widehat{f_{l,k}}$) & $\widehat{B_l}/\mathcal{K}$ \\
queue size per priority ($\widehat{T_{k}}$) & $200/\mathcal{K}$ \\
VNF capacity per service ($\widehat{\mathcal{C}_s}$) & $20$ mbps\\
capacity requirement per request ($\widetilde{\mathcal{C}_r}$) & $\sim \mathcal{U}\{4, 8\}$ mbps\\
bandwidth requirement per request ($\widetilde{\mathcal{B}_r}$) & $\sim \mathcal{U}\{2, 10\}$ mbps\\
traffic burstiness per request ($\widetilde{\mathcal{T}_r}$) & $\sim \mathcal{U}\{1, 4\}$ \\
packet size per request ($\widetilde{\mathcal{H}_r}$) & $1$ \\
\hline
\multicolumn{2}{l}{$x$ is the number of tiers minus the tier number of the node}
\end{tabular}
\label{Tab_SimulationParameters}
\end{center}
\end{table}

\begin{table}[t!]
\caption{Training Configuration.}
\begin{center}
\begin{tabular}{|c|c|}
\hline
\textbf{Parameter} & \textbf{Value} \\
\hline
number of training steps & $10^4$  \\
learning rate & $10^{-4}$ \\
memory size & $5 \times 10^4$ \\
batch size & $32$ \\
discount factor ($\gamma$) & $0.99$ \\
epsilon decrement ($\epsilon'$) & $5 \times 10^{-6}$ \\
epsilon bound ($\widetilde{\epsilon}$) & $5 \times 10^{-2}$ \\
\hline
\end{tabular}
\label{Tab_TrainingConfiguration}
\end{center}
\end{table}

\begin{figure*}[t!]\centering
\includegraphics[width=7.1in]{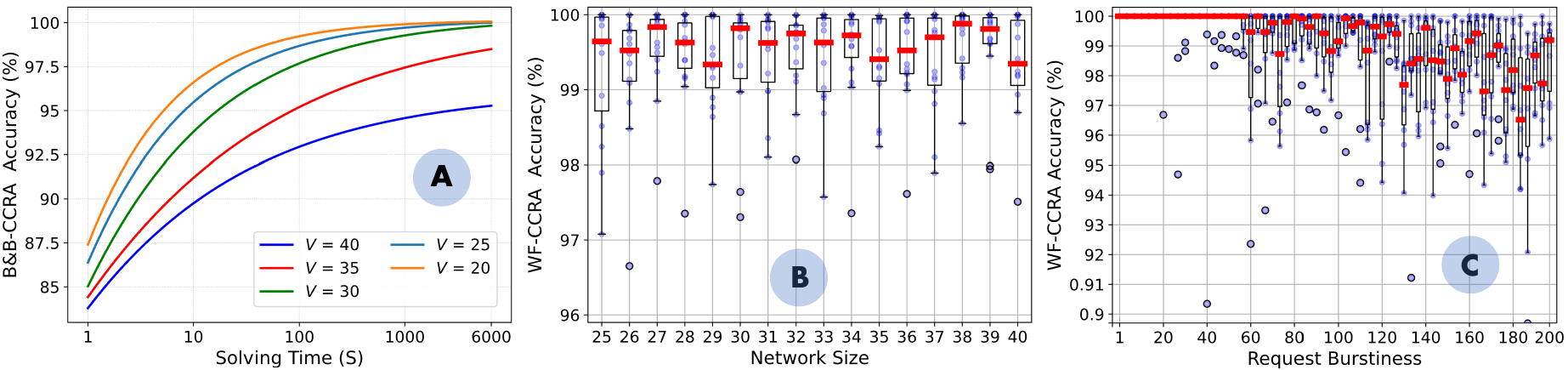}
  \caption{Solution accuracy of A) B\&B-CCRA vs. solving time, B) WF-CCRA vs. network size, and C) WF-CCRA vs. request burstiness. In subfigures A and B, the number of requests is set to $200$, and the number of network nodes in subfigure C is $20$. In subfigure B and C, for each number of nodes or requests, 50 random systems are generated, and the problem is solved using WF-CCRA, with B\&B-CCRA providing the optimal solution. The results of random samples are represented by blue dots, and the aggregated results are represented through boxplots, where red points indicate medians.}
  \label{FIG_bb&wf_accuracy}
\end{figure*}

In this section, the efficiency of the proposed methods is numerically investigated. The system model parameters are listed in Table \ref{Tab_SimulationParameters}, and the configuration of the agents' training procedure is shown in Table \ref{Tab_TrainingConfiguration}. Note that the results are obtained on a computer with 8 processing cores, 16 GB of memory, and a 64-bit operating system.




The accuracy of the B\&B-CCRA and WF-CCRA methods is illustrated in Fig. \ref{FIG_bb&wf_accuracy}. The methods are evaluated based on the accuracy of the solutions they provide. Note that the accuracy of a solution for a scenario named $\eta$ is defined as $1-((\eta-\eta^\star)/\eta^\star)$, where $\eta^\star$ is the scenario's optimal solution, which is obtained by solving it with CPLEX 12.10. In Fig. \ref{FIG_bb&wf_accuracy}.A, the accuracy of B\&B-CCRA is plotted vs. the solving time (in logarithmic scale) for five scenarios with different network sizes. As illustrated, the accuracy of B\&B-CCRA starts at $80\%$ after the first iteration, which is obtained by solving the LP transformation of LiCCRA with CPLEX 12.10 in just a few milliseconds, and increases as the solving time passes, reaching $92\%$ for all samples after 100 seconds. It proves that this method can be easily applied to provide baseline solutions for small and medium size use cases. However, the accuracy growth is slowed by increasing the network size, which is expected given the problem's NP-hardness and complexity. In the two remaining subfigures, the accuracy of WF-CCRA is depicted against the number of requests attending to use system resources, known as request burstiness, and network size. It is evident that regardless of network size, WF-CCRA has an average accuracy greater than $99\%$, implying that it can be used to allocate resources in a near-optimal manner even for large networks. For different numbers of requests, the average accuracy remains significantly high and greater than $96\%$. It does, however, slightly decrease as the number of requests increases, which is the cost of decomplexifying the problem by allocating the resources through isolating requests. Since the decrease is negligible, it is expected that the algorithm is capable of allocating resources efficiently for large numbers of requests.

\begin{figure*}[t!]\centering
\includegraphics[width=7.1in]{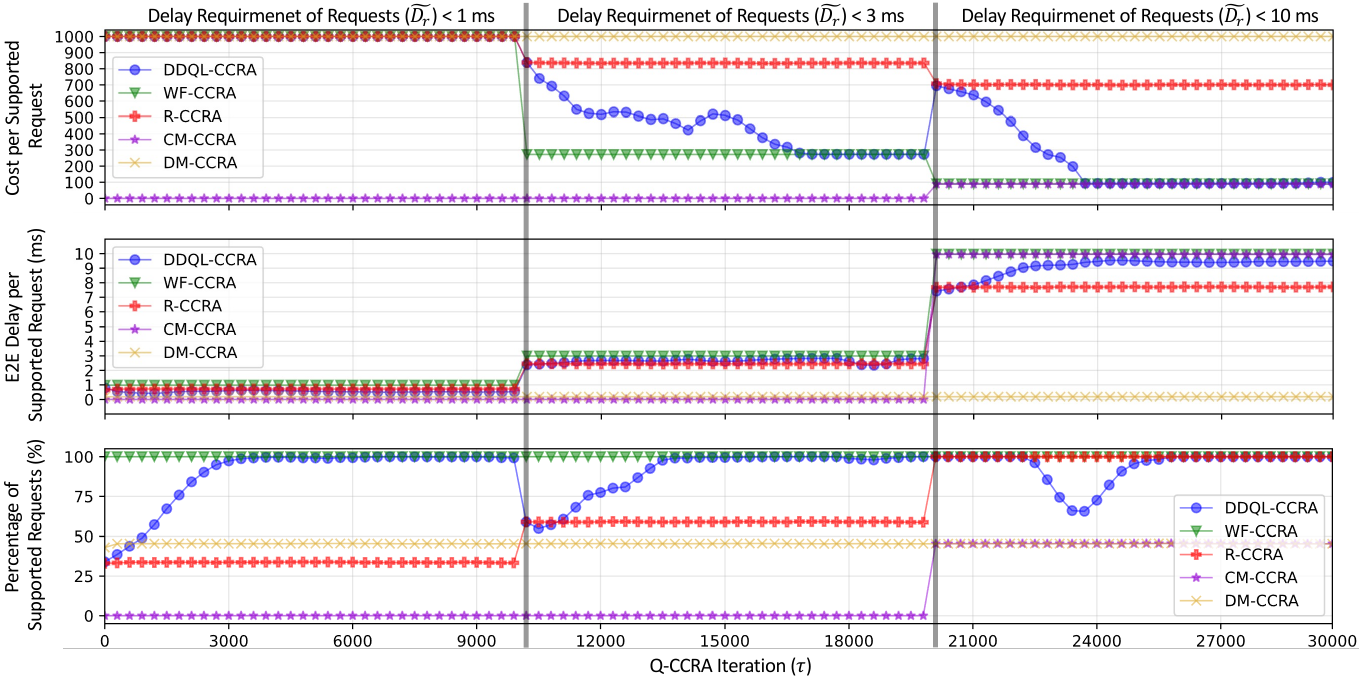}
  \caption{DDQL-CCRA convergence compared with WF-CCRA, R-CCRA, CM-CCRA, and DM-CCRA. The number of requests is $150$, and the number of network nodes is $30$, and all requests can be serviced by a single tier's resources. The results are calculated as a moving average with a window size of $100$ in order to capture trend lines.}
  \label{FIG_DDQLCCRAConvergence}
\end{figure*}

The DDQL-CCRA resource allocation architecture, depicted in Fig. \ref{Fig_DDQLbasedResourceAllocationArchitecture}, is examined in Fig. \ref{FIG_DDQLCCRAConvergence}. In this figure, the mean cost and E2E delay per each supported request, as well as the percentage of supported requests, are plotted against the DDQL-CCRA iteration counter for three scenarios with varying E2E delay requirements. In order to supplement the analysis, this figure additionally includes the outcomes of WF-CCRA in parallel to R-, CM-, and DM-CCRA. In R-CCRA, all allocations are determined at random, but in CM- and DM-CCRA, allocations are made to minimize cost and delay, respectively, without considering other constraints. Note that in order to implement DDQL-CCRA, we deployed the DDQL-CCRA resource allocation architecture on all edge devices (the entry nodes).

When $\widetilde{D_r}$ is less than $1$ ms, the only feasible solution is to assign all requests to the most costly nodes of the first tier. Consequently, the mean cost for all techniques is high, with the exception of CM-CCRA, which attempts to fit all requests into one of the third-tier nodes with the lowest cost, resulting in the inability to support any request and the mean cost of $0$. Since the mean delay for all nodes in the first tier is too low, the average delay per each supported request for all methods excluding CM-CCRA is less than $1$ ms and similar. However, the supported request rate is entirely different for each method. R-CCRA, which assigns nodes evenly to requests, places a third of requests on the first tier, therefore its rate is approximately $33\%$. DM-CCRA selects the node with the shortest E2E delay; hence, its support rate is the number of requests that can be serviced by a single node in the first tier, which is approximately $45\%$. Given that DDQL-CCRA employs the $\epsilon$-greedy technique, it also generates random results during the initial learning iterations. However, as the learning progresses, it receives the reward based on end-user responses and begins to place more and more requests on the first tier until it reaches the near-optimal solutions supplied by WF-CCRA.

When the E2E delay requirement threshold is changed to $3$ ms, both the first and second tier nodes can be occupied to support requests. Since DM- and CM-CCRA always select a node in the first and third tiers, respectively, their outcomes are identical to those of the preceding scenario. R-CCRA doubles the percentage of supported requests because it randomly assigns $66\%$ of requests to the first and second tiers. In addition, its mean delay is slightly smaller than that of WF- and DDQL-CCRA since it utilizes the first tier nodes more than these two cost-effective approaches. Note that the difference is negligible, as the delay of nodes in the first tier is vanishingly small and cannot significantly affect the mean delay. In contrast, when DDQL-CCRA identifies a changing need (lines 35 and 36 of Algorithm \ref{Alg_DDQLCCRA}), it restarts the learning process and enables the $\epsilon$-greedy technique. Therefore, it begins anew with random results and optimizes allocation by fitting as many requests as feasible into the second-tier nodes in ascending cost order. Also in this scenario, it can be observed that the learning technique yields near-optimal efficiency outcomes.

The final scenario is eliminating the delay requirement and releasing the entire infrastructure to serve requests. In this case, although the results for DM-CCRA are identical, the support rate for CM-CCRA is approximately $50\%$, indicating that the node with the lowest cost can service approximately $50\%$ of requests. Similar to the prior scenario, the outcomes of R-CCRA are enhanced. Now it can support all requests, but its mean cost and delay are not optimal because it consumes the resources of all tiers equally. Similarly, the trend for DDQL-CCRA is the same. As soon as it senses a change in requirements, it begins to randomly assign resources, recognizing that requests should be sent as much as possible to the core clouds. It initially determines that the node with the lowest cost yields the best outcome. Therefore, it places all requests on a single node, thereby reducing the number of supported requests and enhancing the mean delay and cost. Subsequently, the reward of this allocation begins to decline as certain requests cannot be supported, the value of dispersing requests throughout the third tier increases progressively, and the optimal policy leads to an increase in the support rate, coupled with a reduction in the mean cost and delay. In Fig. \ref{FIG_DDQLCCRAConvergence}, it is evident that the DDQL-CCRA approach in partially known systems can lead to near-optimal solutions obtained when those systems are fully known. 



\begin{figure*}[t!]\centering
\includegraphics[width=7.1in]{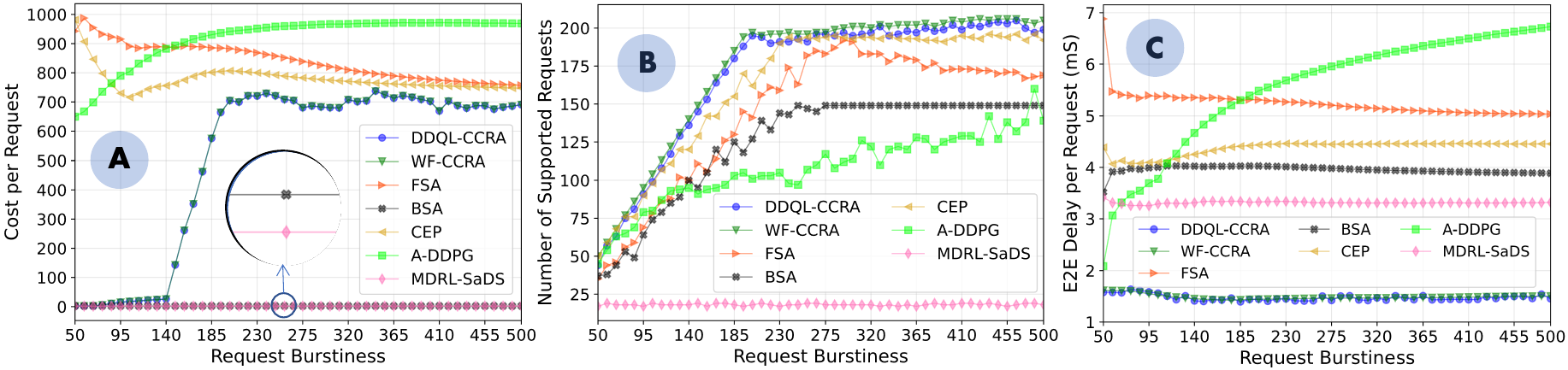}
  \caption{A) Mean cost of each supported request, B) number of supported requests, and C) mean E2E delay of each supported request for DDQL-CCRA, WF-CCRA, FSA \& BSA \cite{nguyen_virtual_2023}, CEP \cite{liu_cost-efficient_2023}, A-DDPG \cite{he_leveraging_2023}, and MDRL-SaDS \cite{xuan_multi-agent_2023} vs. request burstiness. The delay requirement of requests is $10$ ms, and the number of network nodes is $9$. The results are calculated as a moving average with a window size of $100$, where each sample is the average of $50$ arbitrary systems. 
}
  \label{FIG_changing_request_burst}
\end{figure*}

\begin{figure*}[t!]\centering
\includegraphics[width=7.1in]{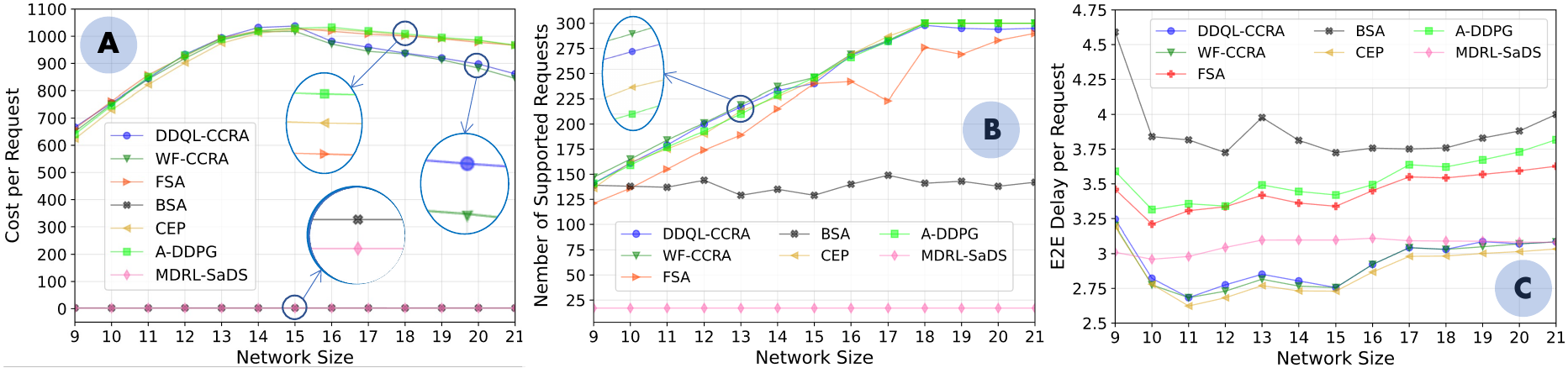}
  \caption{A) Mean cost of each supported request, B) number of supported requests, and C) mean E2E delay of each supported request for DDQL-CCRA, WF-CCRA, FSA \cite{nguyen_virtual_2023}, BSA \cite{nguyen_virtual_2023}, CEP \cite{liu_cost-efficient_2023}, A-DDPG \cite{he_leveraging_2023}, and MDRL-SaDS \cite{xuan_multi-agent_2023} vs. network size. In this scenario, the first four nodes are added to the first tier, followed by the second four nodes to the second tier, and then the last four nodes to the third tier. The delay requirement is $10$ ms, and there are a total of $300$ requests. The results are calculated as a moving average with a window size of $20$, where each sample is the average of $50$ arbitrary systems.}
  \label{FIG_changing_network_size}
\end{figure*}

In Fig. \ref{FIG_changing_request_burst}, DDQL-CCRA is investigated with regard to request burstiness. The mean cost and E2E delay of each supported request are depicted in Fig. \ref{FIG_changing_request_burst}.A and Fig. \ref{FIG_changing_request_burst}.C respectively, whereas Fig. \ref{FIG_changing_request_burst}.B illustrates the number of supported requests. In this figure, the results of DDQL-CCRA are compared with those of WF-CCRA, FSA \cite{nguyen_virtual_2023}, BSA \cite{nguyen_virtual_2023}, CEP \cite{liu_cost-efficient_2023}, A-DDPG \cite{he_leveraging_2023}, and MDRL-SaDS \cite{xuan_multi-agent_2023}. FSA is a heuristic algorithm that randomly assigns resources to requests in descending order of their required computing capacity. BSA is a similar method that assigns resources in descending order of their remaining capacity to the sorted requests. In CEP, resources are allocated with the aim of minimizing the total cost of links. A-DDPG is an RL method that adjusts the reward for each request to maximize its overall utility. In this solution, utility is defined as the profit of serving the request as a function of its required bandwidth minus the E2E path delay experienced. MDRL-SaDS is another RL technique in which the reward is the computing and networking cost of serving each request divided by the total cost of allocated resources across the infrastructure. This strategy seeks to minimize the cost of allocated resources in relation to their energy consumption.

Evidently, the number of requests supported by FSA is relatively high, as are its mean cost and E2E delay. FSA distributes requests across all tiers, resulting in significant utilization of all resources, a high mean cost, and a high mean E2E delay. Since all links have a similar cost, CEP exhibits comparable performance; however, because it considers the feasibility of links, it achieves slightly better results. A-DDPG, where requests are assigned to nodes with a lower E2E delay, is another costly method. Increasing the number of requests causes second- and first-tier nodes to become occupied and requests to be assigned to the resources of other tires, thereby increasing the E2E delay. In BSA, because the performance metric is the remaining capacity of computing nodes and the nodes are ordered from low capacity to high capacity across the tiers, it occupies the nodes from the cloud to the edge, resulting in outcomes with very low cost and moderate delay. A-DDPG and BSA cannot support a substantial number of requests because the feasibility of links is not explicitly evaluated. MDRL-SaDS is the most inefficient method by which requests are routed to the node with the lowest cost. Therefore, the number of supported requests is proportional to the node's capacity and delay. The behavior of the two remaining methods, DDQL- and WF-CCRA, is comparable. They support requests by initially assigning third-tier nodes. Then, once this tier is occupied, they proceed to occupy the second tier, resulting in an exponential cost increase. The mean cost converges to a fixed value when all resources are occupied and the first tier is in use. This approach results in a very low E2E delay because it assigns request priorities based on their delay requirements (unlike other approaches, which are unaware of ATS queues). DDQL-CCRA can provide near-optimal solutions regardless of the number of requests received, as demonstrated. \vspace{1.1em}

The final figure compares the proposed approaches to the approaches depicted in the preceding figure for various network sizes. In this scenario, if $\mathcal{V} \leq 15$, the first-tier network has a very high capacity, whereas if $\mathcal{V} > 15$, there are sufficient resources to fulfill all requests. Therefore, CEP and A-DDPG, which focus on minimizing the cost of allocated links and E2E delay respectively, as well as FSA, which allocates resources randomly, can support a large number of requests despite the high cost of allocations. Since the capacity of the low-cost tier for BSA and MDRL-SaDS is not excessive (and the capacity ratio of this tier to the others is less than the previous figure), the request support rate is unpromising despite the low cost. When the infrastructure is full ($\mathcal{V} \leq 15$), the results of WF- and DDQL-CCRA are comparable to those of other algorithms. However, when there are more resources ($\mathcal{V} > 15$), these two approaches move requests to low-cost resources, thereby reducing the total cost of allocations. When it comes to E2E delay, even though the results are similar for all methods, by adding a node to the initial network, the resources of the first tier are extended and more requests can be supported with smaller delays, resulting in a sudden decrease for $\mathcal{V} = 10$. By adding more nodes, however, more resources are added to the other tires, and FSA (which allocates resources randomly), BSA (which assigns resources in descending order of their remaining capacity), and A-DDPG (which tries to maximize the overall utility) migrate requests to the lower-cost tiers, resulting in a slight increase in E2E delay. Despite the increase in WF- and DDQL-CCRA techniques, their outcome is the lowest E2E delay because they manage priority queues according to the delay requirements of requests.



\section{Conclusion}\label{Sec_Conclusion}
In this paper, the joint problem of communication and computing resource allocation comprising VNF placement and assignment, traffic prioritization, and path selection considering capacity and delay constraints was studied. The primary objective was to minimize the total cost of allocations. We initially formulated the problem as a MINLP model, and used a method, named B\&B-CCRA, to solve it optimally. Then, due to the complexity of B\&B-CCRA, a WF-based approach was developed to find near-optimal solutions in a timely manner. These two methods can be utilized to solve the problem when the system is fully known. However, for scenarios wherein there is no request-specific information, a DDQL-based architecture was presented that yields near-optimal solutions. The efficiency of the proposed methods was demonstrated by numerical results.

As potential future work, we intend to address the problem by accounting for more dynamic environments in which end-users are mobile and all of their needs are subject to change. In addition, the proposed methods could be supplemented by taking into account dynamic infrastructure resources, in which the cost of resources (such as their energy consumption) or their availability can fluctuate over time. In such highly dynamic scenarios, we intend to enhance the proposed DDQL-based method with Continual Learning in order to reduce the adaptation time required to adjust agents after each change. Another possible research direction is to extend the problem to include radio domain resources (such as power control, channel assignments, rate control, and relay selection in multi-hop scenarios), thereby providing end-user-to-end-user resource allocations. Furthermore, we intend to improve the proposed method for allocating resources to VNF chains rather than individual VNFs.

\section*{Acknowledgment}
This research work is partially supported by the Business Finland 6Bridge 6Core project under Grant No. 8410/31/2022, the Academy of Finland IDEA-MILL project under Grant No. 352428, the European Union’s Horizon 2020 ICT Cloud Computing program under the ACCORDION project with grant agreement No. 871793, and the Academy of Finland 6G Flagship program under Grant No. 346208. 

\bibliographystyle{IEEEtran}
\bibliography{IEEEabrv,main}

\begin{thebibliography}{10}
\providecommand{\url}[1]{#1}
\csname url@rmstyle\endcsname
\providecommand{\newblock}{\relax}
\providecommand{\bibinfo}[2]{#2}
\providecommand\BIBentrySTDinterwordspacing{\spaceskip=0pt\relax}
\providecommand\BIBentryALTinterwordstretchfactor{4}
\providecommand\BIBentryALTinterwordspacing{\spaceskip=\fontdimen2\font plus
\BIBentryALTinterwordstretchfactor\fontdimen3\font minus
  \fontdimen4\font\relax}
\providecommand\BIBforeignlanguage[2]{{%
\expandafter\ifx\csname l@#1\endcsname\relax
\typeout{** WARNING: IEEEtran.bst: No hyphenation pattern has been}%
\typeout{** loaded for the language `#1'. Using the pattern for}%
\typeout{** the default language instead.}%
\else
\language=\csname l@#1\endcsname
\fi
#2}}

\bibitem{taleb_towards_nodate}
T.~Taleb, A.~Boudi, L.~Rosa, L.~Cordeiro, T.~Theodoropoulos, K.~Tsepes,
  P.~Dazzi, A.~Protopsaltis, and R.~Li, ``Towards {Supporting} {XR} {Services}:
  {Architecture} and {Enablers},'' \emph{IEEE Internet of Things Journal}.

\bibitem{corneo_surrounded_2021}
L.~Corneo, M.~Eder, Mohan, \emph{et~al.}, ``Surrounded by the {Clouds}: {A}
  {Comprehensive} {Cloud} {Reachability} {Study},'' in \emph{Proceedings of the
  {Web} {Conference} 2021}, ser. {WWW} '21.\hskip 1em plus 0.5em minus
  0.4em\relax New York, NY, USA: Association for Computing Machinery, Apr.
  2021, pp. 295--304.

\bibitem{yang_urllc_2021}
X.~Yang, Z.~Zho, and B.~Huang, ``{URLLC} {Key} {Technologies} and
  {Standardization} for {6G} {Power} {Internet} of {Things},'' \emph{IEEE
  Communications Standards Magazine}, vol.~5, no.~2, pp. 52--59, June 2021.

\bibitem{taleb_multi-domain_2019}
T.~Taleb, I.~Afolabi, K.~Samdanis, and F.~Z. Yousaf, ``On {Multi}-{Domain}
  {Network} {Slicing} {Orchestration} {Architecture} and {Federated} {Resource}
  {Control},'' \emph{IEEE Network}, vol.~33, no.~5, pp. 242--252, Sept. 2019.

\bibitem{taleb_cdn_2020}
T.~Taleb, P.~A. Frangoudis, I.~Benkacem, and A.~Ksentini, ``{CDN} {Slicing}
  over a {Multi}-{Domain} {Edge} {Cloud},'' \emph{IEEE Transactions on Mobile
  Computing}, vol.~19, no.~9, pp. 2010--2027, Sept. 2020.

\bibitem{li_cognitive_2021}
Y.~Li, J.~Huang, Q.~Sun, T.~Sun, and S.~Wang, ``Cognitive {Service}
  {Architecture} for {6G} {Core} {Network},'' \emph{IEEE Transactions on
  Industrial Informatics}, vol.~17, no.~10, pp. 7193--7203, Oct. 2021.

\bibitem{emu_latency_2020}
M.~Emu, P.~Yan, and S.~Choudhury, ``Latency {Aware} {VNF} {Deployment} at
  {Edge} {Devices} for {IoT} {Services}: {An} {Artificial} {Neural} {Network}
  {Based} {Approach},'' in \emph{2020 {IEEE} {International} {Conference} on
  {Communications} {Workshops} ({ICC} {Workshops})}, June 2020, pp. 1--6, iSSN:
  2474-9133.

\bibitem{vasilakos_towards_2021}
X.~Vasilakos, M.~Bunyakitanon, R.~Nejabati, and D.~Simeonidou, ``Towards
  {Low}-latent \& {Load}-balanced {VNF} {Placement} with {Hierarchical}
  {Reinforcement} {Learning},'' in \emph{2021 {IEEE} {International}
  {Mediterranean} {Conference} on {Communications} and {Networking}
  ({MeditCom})}, Sept. 2021, pp. 162--167.

\bibitem{sami_demand-driven_2021}
H.~Sami, A.~Mourad, H.~Otrok, and J.~Bentahar, ``Demand-{Driven} {Deep}
  {Reinforcement} {Learning} for {Scalable} {Fog} and {Service} {Placement},''
  \emph{IEEE Transactions on Services Computing}, pp. 1--1, 2021.

\bibitem{liu_cost-efficient_2023}
M.~Liu and S.~B. Alias, ``Cost-{Efficient} {Virtual} {Network} {Function}
  {Placement} in an {Industrial} {Edge} {System}: {A} {Proposed} {Method},''
  \emph{IEEE Systems, Man, and Cybernetics Magazine}, vol.~9, no.~1, pp.
  10--17, Jan. 2023.

\bibitem{he_leveraging_2023}
N.~He, S.~Yang, F.~Li, S.~Trajanovski, L.~Zhu, Y.~Wang, and X.~Fu, ``Leveraging
  {Deep} {Reinforcement} {Learning} {With} {Attention} {Mechanism} for
  {Virtual} {Network} {Function} {Placement} and {Routing},'' \emph{IEEE
  Transactions on Parallel and Distributed Systems}, vol.~34, no.~4, pp.
  1186--1201, Apr. 2023.

\bibitem{iwamoto_optimal_2023}
M.~Iwamoto, A.~Suzuki, and M.~Kobayashi, ``Optimal {VNF} {Scheduling} for
  {Minimizing} {Duration} of {QoS} {Degradation},'' in \emph{2023 {IEEE} 20th
  {Consumer} {Communications} \& {Networking} {Conference} ({CCNC})}, Jan.
  2023, pp. 855--858, iSSN: 2331-9860.

\bibitem{nguyen_virtual_2023}
D.~H.~P. Nguyen, Y.-H. Lien, B.-H. Liu, S.-I. Chu, and T.~N. Nguyen, ``Virtual
  {Network} {Function} {Placement} for {Serving} {Weighted} {Services} in
  {NFV}-{Enabled} {Networks},'' \emph{IEEE Systems Journal}, pp. 1--12, 2023.

\bibitem{xuan_multi-agent_2023}
H.~Xuan, Y.~Zhou, X.~Zhao, and Z.~Liu, ``\BIBforeignlanguage{en}{Multi-agent
  deep reinforcement learning algorithm with self-adaption division strategy
  for {VNF}-{SC} deployment in {SDN}/{NFV}-{Enabled} {Networks}},''
  \emph{\BIBforeignlanguage{en}{Applied Soft Computing}}, vol. 138, p. 110189,
  May 2023.

\bibitem{miyamura_joint_2023}
T.~Miyamura and A.~Misawa, ``\BIBforeignlanguage{en}{Joint optimization of
  optical path provisioning and {VNF} placement in {vCDN}},''
  \emph{\BIBforeignlanguage{en}{Optical Switching and Networking}}, vol.~49, p.
  100740, May 2023.

\bibitem{yang_online_nodate}
C.~Yang, B.~Hu, Y.~Feng, H.~Huang, H.~Lai, and J.~Tan,
  ``\BIBforeignlanguage{en}{An online service function chain orchestration
  method for profit maximization in edge computing networks},''
  \emph{\BIBforeignlanguage{en}{Engineering Reports}}, vol. n/a, no. n/a, p.
  e12653.

\bibitem{kuo_deploying_2018}
T.-W. Kuo, B.-H. Liou, K.~C.-J. Lin, and M.-J. Tsai, ``Deploying {Chains} of
  {Virtual} {Network} {Functions}: {On} the {Relation} {Between} {Link} and
  {Server} {Usage},'' \emph{IEEE/ACM Transactions on Networking}, vol.~26,
  no.~4, pp. 1562--1576, Aug. 2018.

\bibitem{mada_latency-aware_2020}
B.~E. Mada, M.~Bagaa, T.~Tale, and H.~Flinck, ``Latency-aware {Service}
  {Placement} and {Live} {Migrations} in {5G} and {Beyond} {Mobile}
  {Systems},'' in \emph{{ICC} 2020 - 2020 {IEEE} {International} {Conference}
  on {Communications} ({ICC})}, June 2020, pp. 1--6, iSSN: 1938-1883.

\bibitem{zhang_adaptive_2019}
Q.~Zhang, F.~Liu, and C.~Zeng, ``Adaptive {Interference}-{Aware} {VNF}
  {Placement} for {Service}-{Customized} {5G} {Network} {Slices},'' in
  \emph{{IEEE} {INFOCOM} 2019 - {IEEE} {Conference} on {Computer}
  {Communications}}, Apr. 2019, pp. 2449--2457, iSSN: 2641-9874.

\bibitem{yuan_toward_2020}
Q.~Yuan, X.~Ji, H.~Tang, and W.~You, ``Toward {Latency}-{Optimal} {Placement}
  and {Autoscaling} of {Monitoring} {Functions} in {MEC},'' \emph{IEEE Access},
  vol.~8, pp. 41\,649--41\,658, 2020.

\bibitem{gao_cost-efficient_2020}
T.~Gao, X.~Li, Y.~Wu, W.~Zou, S.~Huang, M.~Tornatore, and B.~Mukherjee,
  ``Cost-{Efficient} {VNF} {Placement} and {Scheduling} in {Public} {Cloud}
  {Networks},'' \emph{IEEE Transactions on Communications}, vol.~68, no.~8, pp.
  4946--4959, Aug. 2020.

\bibitem{alwis_survey_2021}
C.~D. Alwis, A.~Kalla, Q.-V. Pham, P.~Kumar, K.~Dev, W.-J. Hwang, and
  M.~Liyanage, ``Survey on {6G} {Frontiers}: {Trends}, {Applications},
  {Requirements}, {Technologies} and {Future} {Research},'' \emph{IEEE Open
  Journal of the Communications Society}, vol.~2, pp. 836--886, 2021.

\bibitem{shokrnezhad_near-optimal_2022}
M.~Shokrnezhad and T.~Taleb, ``Near-optimal {Cloud}-{Network} {Integrated}
  {Resource} {Allocation} for {Latency}-{Sensitive} {B5G},'' in \emph{2022
  {IEEE} {Global} {Communications} {Conference} ({GLOBECOM})}, Rio De Janeiro,
  Brazil, Dec. 2022.

\bibitem{lei_energy-saving_2021}
J.~Lei, S.~Deng, Z.~Lu, Y.~He, and X.~Gao,
  ``\BIBforeignlanguage{en}{Energy-saving traffic scheduling in backbone
  networks with software-defined networks},''
  \emph{\BIBforeignlanguage{en}{Cluster Computing}}, vol.~24, no.~1, pp.
  279--292, Mar. 2021.

\bibitem{specht_urgency-based_2016}
J.~Specht and S.~Samii, ``Urgency-{Based} {Scheduler} for {Time}-{Sensitive}
  {Switched} {Ethernet} {Networks},'' in \emph{2016 28th {Euromicro}
  {Conference} on {Real}-{Time} {Systems} ({ECRTS})}, July 2016, pp. 75--85,
  iSSN: 2159-3833.

\bibitem{arshad_utilizing_2022}
U.~Arshad, M.~Aleem, G.~Srivastava, and J.~C.-W. Lin,
  ``\BIBforeignlanguage{en}{Utilizing power consumption and {SLA} violations
  using dynamic {VM} consolidation in cloud data centers},''
  \emph{\BIBforeignlanguage{en}{Renewable and Sustainable Energy Reviews}},
  vol. 167, p. 112782, Oct. 2022.

\bibitem{kianpisheh_survey_nodate}
S.~Kianpisheh and T.~Taleb, ``A {Survey} on {In}-network {Computing}:
  {Programmable} {Data} {Plane} {And} {Technology} {Specific} {Applications},''
  \emph{IEEE Communications Surveys and Tutorials (COMST)}.

\bibitem{bhat_6g_2021}
J.~R. Bhat and S.~A. Alqahtani, ``{6G} {Ecosystem}: {Current} {Status} and
  {Future} {Perspective},'' \emph{IEEE Access}, vol.~9, pp. 43\,134--43\,167,
  2021.

\bibitem{giordani_toward_2020}
M.~Giordani, M.~Polese, M.~Mezzavilla, S.~Rangan, and M.~Zorzi, ``Toward {6G}
  {Networks}: {Use} {Cases} and {Technologies},'' \emph{IEEE Communications
  Magazine}, vol.~58, no.~3, pp. 55--61, Mar. 2020.

\bibitem{gustavsson_implementation_2021}
U.~Gustavsson, P.~Frenger, C.~Fager, T.~Eriksson, H.~Zirath, F.~Dielacher,
  C.~Studer, A.~Pärssinen, R.~Correia, J.~N. Matos, D.~Belo, and N.~B.
  Carvalho, ``Implementation {Challenges} and {Opportunities} in {Beyond}-{5G}
  and {6G} {Communication},'' \emph{IEEE Journal of Microwaves}, vol.~1, no.~1,
  pp. 86--100, Jan. 2021.

\bibitem{mahmood_machine_2021}
N.~H. Mahmood, S.~Böcker, Moerman, \emph{et~al.},
  ``\BIBforeignlanguage{en}{Machine type communications: key drivers and
  enablers towards the {6G} era},'' \emph{\BIBforeignlanguage{en}{EURASIP
  Journal on Wireless Communications and Networking}}, vol. 2021, no.~1, p.
  134, June 2021.

\bibitem{yu_deep_nodate}
H.~Yu, C.~Wang, T.~Taleb, and J.~Zhang, ``Deep {Reinforcement} {Learning} based
  {Deterministic} {Routing} and {Scheduling} for {Mixed}-{Criticality}
  {Flows},'' \emph{IEEE Transactions on Industrial Informatics}.

\bibitem{guo_probabilistic-assured_nodate}
Q.~Guo, R.~Gu, H.~Yu, T.~Taleb, and Y.~Ji, ``Probabilistic-{Assured} {Resource}
  {Provisioning} with {Customizable} {Hybrid} {Isolation} for {Vertical}
  {Industrial} {Slicing},'' \emph{IEEE Transactions on Network and Service
  Management (TNSM)}.

\bibitem{taleb_orchestrating_2019}
T.~Taleb, I.~Afolabi, and M.~Bagaa, ``Orchestrating {5G} {Network} {Slices} to
  {Support} {Industrial} {Internet} and to {Shape} {Next}-{Generation} {Smart}
  {Factories},'' \emph{IEEE Network}, vol.~33, no.~4, pp. 146--154, July 2019.

\bibitem{kellerer_multidimensional_2004}
H.~Kellerer, U.~Pferschy, and D.~Pisinger,
  ``\BIBforeignlanguage{en}{Multidimensional {Knapsack} {Problems}},'' in
  \emph{\BIBforeignlanguage{en}{Knapsack {Problems}}}, H.~Kellerer,
  U.~Pferschy, and D.~Pisinger, Eds.\hskip 1em plus 0.5em minus 0.4em\relax
  Berlin, Heidelberg: Springer, 2004, pp. 235--283.

\bibitem{pataki_basis_2010}
G.~Pataki, M.~Tural, and E.~B. Wong, ``Basis {Reduction} and the {Complexity}
  of {Branch}-and-{Bound},'' in \emph{Proceedings of the 2010 {Annual}
  {ACM}-{SIAM} {Symposium} on {Discrete} {Algorithms} ({SODA})}, ser.
  Proceedings.\hskip 1em plus 0.5em minus 0.4em\relax Society for Industrial
  and Applied Mathematics, Jan. 2010, pp. 1254--1261.

\bibitem{hasselt_deep_2016}
H.~v. Hasselt, A.~Guez, and D.~Silver, ``\BIBforeignlanguage{en}{Deep
  {Reinforcement} {Learning} with {Double} {Q}-{Learning}},''
  \emph{\BIBforeignlanguage{en}{Proceedings of the AAAI Conference on
  Artificial Intelligence}}, vol.~30, no.~1, Mar. 2016, number: 1.

\bibitem{watkins_technical_1992}
C.~J. C.~H. Watkins and P.~Dayan, ``\BIBforeignlanguage{en}{Technical
  {Note}},'' in \emph{\BIBforeignlanguage{en}{Reinforcement {Learning}}}, ser.
  The {Springer} {International} {Series} in {Engineering} and {Computer}
  {Science}, R.~S. Sutton, Ed.\hskip 1em plus 0.5em minus 0.4em\relax Boston,
  MA: Springer US, 1992, pp. 55--68.

\bibitem{mnih_human-level_2015}
V.~Mnih, K.~Kavukcuoglu, D.~Silver, A.~A. Rusu, J.~Veness, M.~G. Bellemare,
  A.~Graves, M.~Riedmiller, A.~K. Fidjeland, G.~Ostrovski, S.~Petersen,
  C.~Beattie, A.~Sadik, I.~Antonoglou, H.~King, D.~Kumaran, D.~Wierstra,
  S.~Legg, and D.~Hassabis, ``\BIBforeignlanguage{en}{Human-level control
  through deep reinforcement learning},''
  \emph{\BIBforeignlanguage{en}{Nature}}, vol. 518, no. 7540, pp. 529--533,
  Feb. 2015.

\bibitem{kingma2014adam}
D.~P. Kingma and J.~Ba, ``Adam: A method for stochastic optimization,''
  \emph{arXiv preprint arXiv:1412.6980}, 2014.

\bibitem{watkins1992q}
C.~J. Watkins and P.~Dayan, ``Q-learning,'' \emph{Machine learning}, vol.~8,
  pp. 279--292, 1992.

\bibitem{nguyen_digital_2021}
H.~X. Nguyen, R.~Trestian, D.~To, and M.~Tatipamula, ``Digital {Twin} for {5G}
  and {Beyond},'' \emph{IEEE Communications Magazine}, vol.~59, no.~2, pp.
  10--15, Feb. 2021.

\end{thebibliography}

\vspace{1em}
\begin{wrapfigure}{l}{0.1\textwidth}
    \centering
    \vspace{-1em}
    \includegraphics[width=0.1\textwidth]{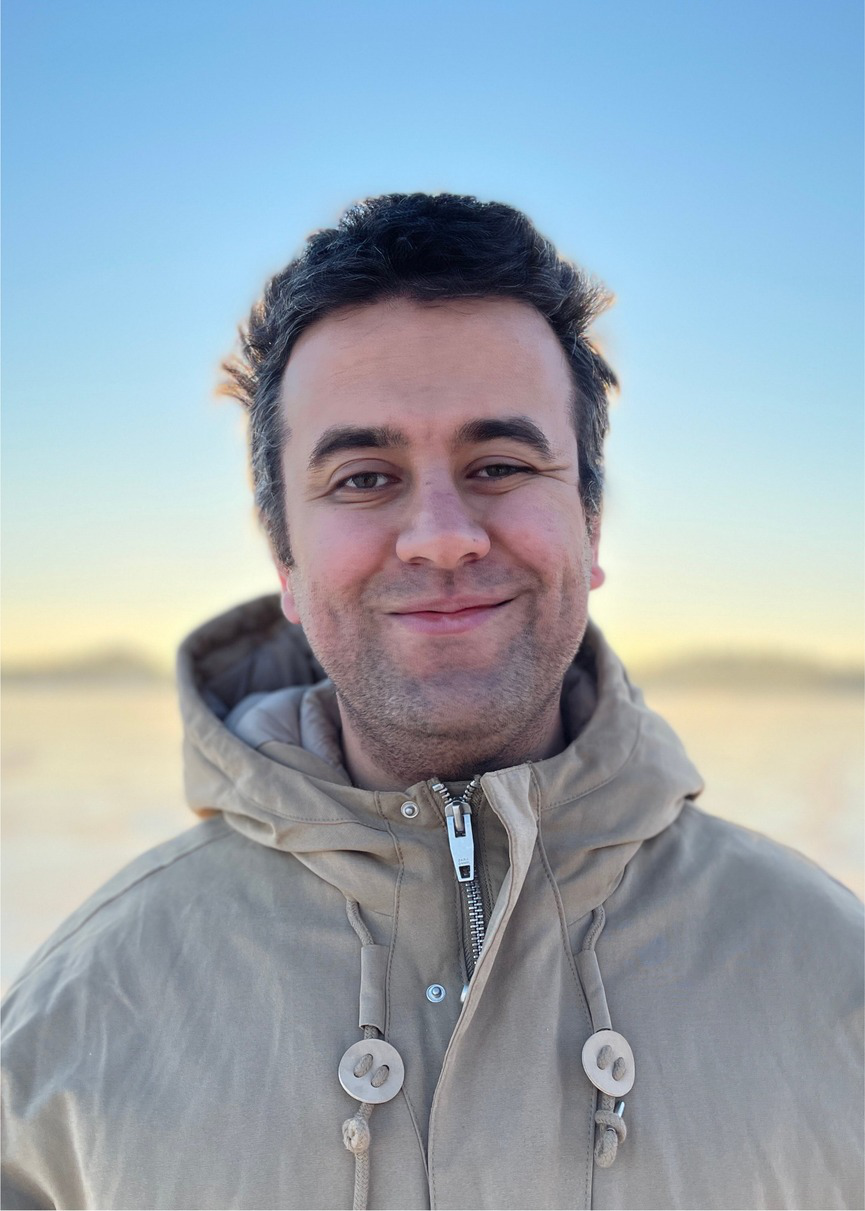}
    \vspace{0em}
\end{wrapfigure}
\noindent \footnotesize \textbf{Masoud Shokrnezhad} received his B.Sc. degree in information technology from Shahid Madani University of Azerbaijan, Tabriz, Iran, and his M.Sc. and Ph.D. degrees (as a bright talent) in computer networks from Amirkabr University of Technology (Tehran Polytechnic), Tehran, Iran, in 2011, 2013, and 2019, respectively. He is currently a postdoctoral researcher with the Center of Wireless Communications, University of Oulu, Oulu, Finland. Between June 2021 and December 2021, he was a postdoctoral researcher at the School of Electrical Engineering, Aalto University, Espoo, Finland. Prior to that, he had been working as a senior system designer and engineer with FavaPars and Pouya Cloud Technology, Tehran, Iran, since 2013. He has been engaged in many national, international, and European projects working on designing and developing computing and networking frameworks, and has been directly co-managing a startup focusing on developing SDWAN solutions for B2B use cases. His research interests lie in the fields of artificial intelligence and machine learning, operations research and optimization theory, semantic networking and communications, as well as resource allocation. In addition to this, he is a full-stack developer who has extensive practical experience working with JS-based technologies.

\vspace{1em}
\begin{wrapfigure}{l}{0.1\textwidth}
    \centering
    \vspace{-1.3em}
    \includegraphics[width=0.1\textwidth]{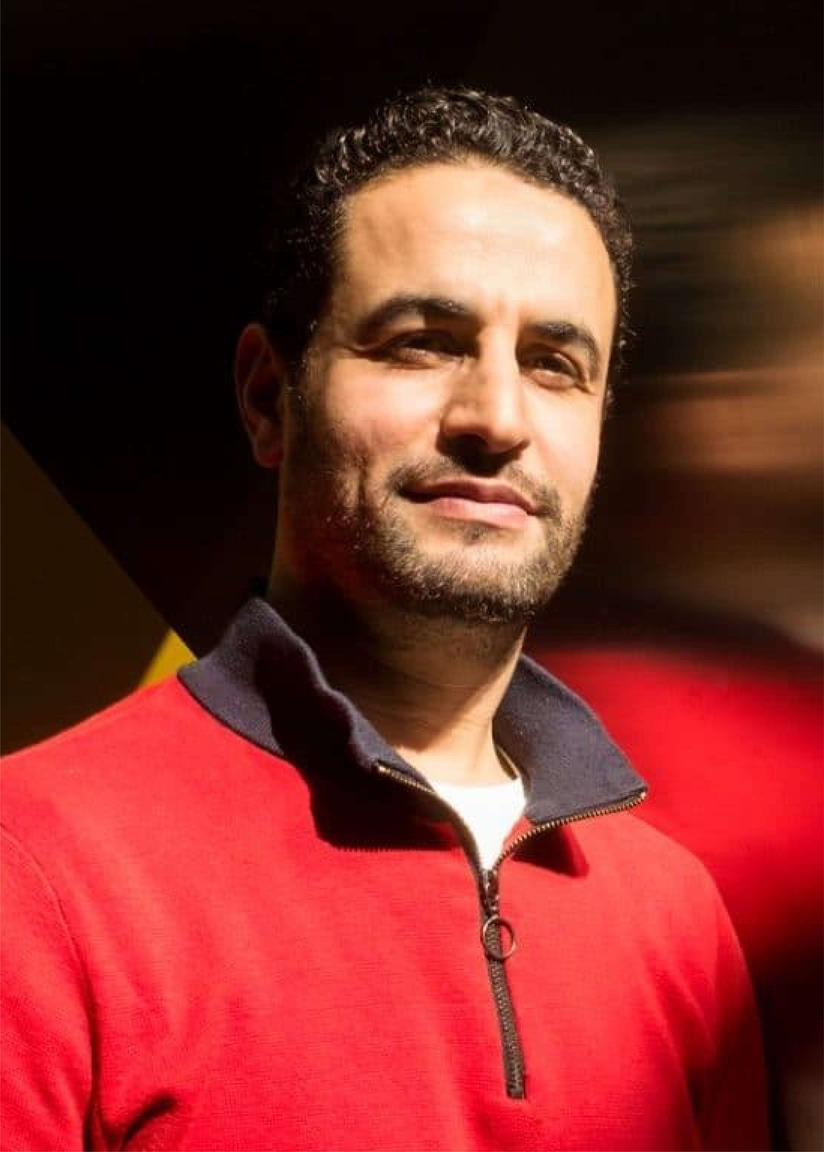}
    \vspace{-2em}
\end{wrapfigure}
\noindent \footnotesize \textbf{Tarik Taleb} (Senior Member, IEEE) received the B.E. degree (with distinction) in information engineering and the M.Sc. and Ph.D. degrees in information sciences from Tohoku University, Sendai, Japan, in 2001, 2003, and 2005, respectively. He is currently a professor with the Center of Wireless Communications, University of Oulu, Oulu, Finland. He is the founder and director of MOSA!C Lab, Espoo, Finland. He has also been directly engaged in the development and standardization of the Evolved Packet System as a member of 3GPP’s System Architecture Working Group 2. His research interests lie in the fields of telco cloud, network softwarization, network slicing, AI-based software-defined security, immersive communications, mobile multimedia streaming, and next-generation mobile networking. Prof. Taleb is the recipient of the 2021 IEEE ComSoc Wireless Communications Technical Committee Recognition Award in December 2021, and the 2017 IEEE ComSoc Communications Software Technical Achievement Award in December 2017 for his outstanding contributions to network softwarization. He is also the co-recipient of the 2017 IEEE Communications Society Fred W. Ellersick Prize in May 2017, and many other awards from Japan.

\vspace{1em}
\begin{wrapfigure}{l}{0.1\textwidth}
    \centering
    \vspace{-1.3em}
    \includegraphics[width=0.1\textwidth]{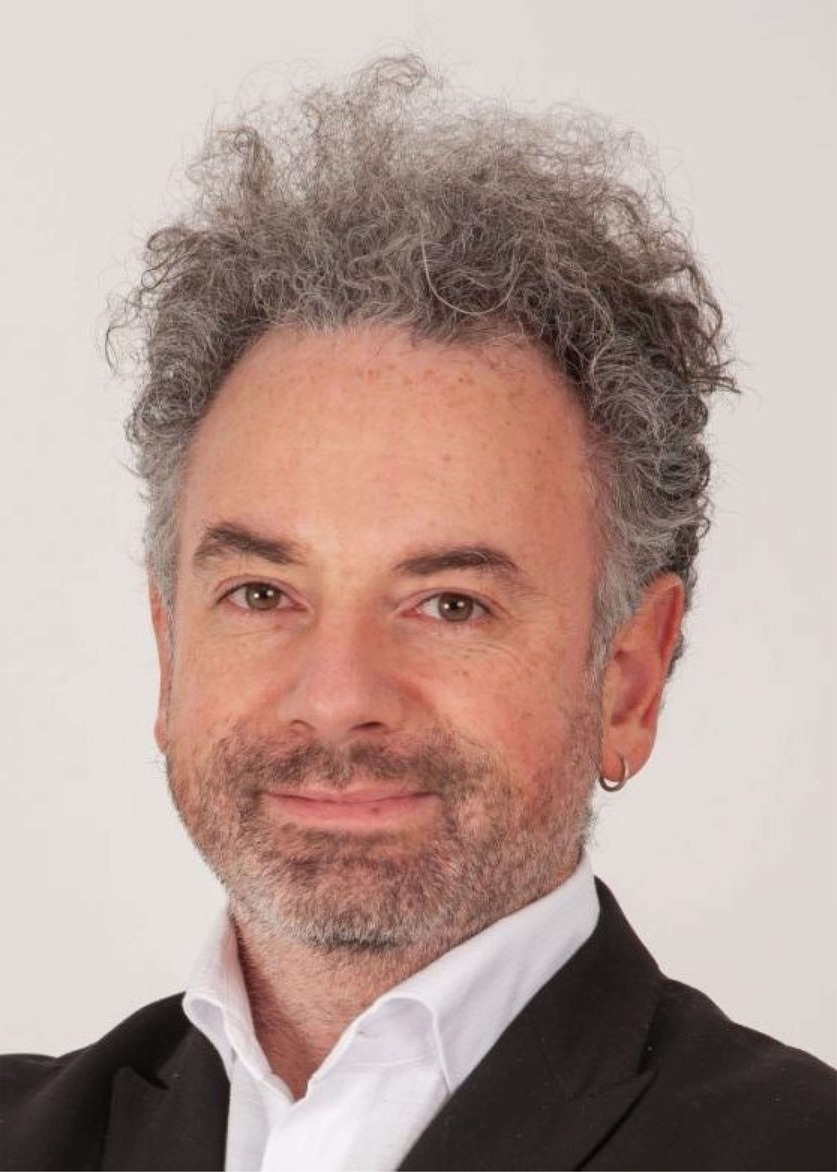}
    \vspace{-2.2em}
\end{wrapfigure}
\noindent \footnotesize \textbf{Patrizio Dazzi} received the B.Sc. degree in computer science and the M.Sc. degree in computer technologies (computer systems and networks) degree from the University of Pisa, Pisa, Italy, in 2003 and 2004, respectively, and the Ph.D. degree in computer science and engineering from the IMT School for Advanced Studies Lucca, Lucca, Italy, in 2008. He is a researcher with a strong interest in high-performance distributed systems, particularly in models and algorithms for the decentralized management of computations and data. He is currently a senior researcher with the University of Pisa, Pisa, Italy, and the co-founder and co-leader of the Pervasive AI Laboratory, a joint initiative of the University of Pisa and the National Research Council of Italy. He has been the Scientific Coordinator of the EU-South Korea H2020 BASMATI project, he served as a Project Coordinator for the EU H2020 ACCORDION project, and he is currently serving as an Innovation Coordinator for the EU H2020 TEACHING project. Dr. Dazzi served as a program committee member at many conferences in the field of large, distributed systems, including IEEE ICDCS, IEEE CLOUD, ACM SoCC, EuroPar, ACM SIGKDD, ACM WSDM, and IEEE HPCS. He is a member of the Executive Committee of the IEEE Technical Committee on Cloud Computing. He has a long-lasting experience as a speaker in international venues.


\end{document}